\documentclass[epjCONF]{svjour}
\usepackage{graphicx}
\usepackage{amssymb}
\usepackage{amsmath}
\usepackage[varg]{txfonts} 
\usepackage[latin1]{inputenc}
\session-title{%
19$^{\textnormal{\footnotesize th}}$ International %
IUPAP Conference on Few-Body Problems in Physics%
}
\begin{document}
\title{%
Scattering of light nuclei
}%
\author{%
S. Quaglioni\inst{1}\fnmsep\thanks{\email{ quaglioni1@llnl.gov }} 
\and %
P. Navr{\'a}til\inst{1} 
\and %
R. Roth\inst{2} 
}
\institute{%
Lawrence Livermore National Laboratory, P.O. Box 808, Livermore, CA 94551, USA
\and %
Institut f\"{u}r Kernphysik, Technische Universit\"{a}t Darmstadt, 64289 Darmstadt, Germany}
\abstract{
The exact treatment of nuclei starting from the constituent nucleons and the fundamental interactions among them has been a long-standing goal in nuclear physics. Above all nuclear scattering and reactions, which require the solution of the many-body quantum-mechanical problem in the continuum, represent an extraordinary theoretical as well as computational challenge for {\em ab initio} approaches.
We present a new {\em ab initio} many-body approach which derives from the combination of the {\em ab initio} no-core shell model with the resonating-group method [4]. By complementing a microscopic cluster technique with the use of realistic interactions, and a microscopic and consistent description of the nucleon clusters, this approach is capable of describing simultaneously both bound and scattering states in light nuclei. We will discuss applications to neutron and proton scattering on $s$- and light $p$-shell nuclei using realistic nucleon-nucleon potentials, and outline the progress toward the treatment of more complex reactions.} 
\maketitle
%
%
%
\section{Introduction}
\label{QuaglioniS_intro}

To understand the evolution of the Universe, we need to understand nuclear reactions. 
Indeed, low-energy fusion reactions represent the primary energy-generation mechanism in stars, help determining the course of stellar evolution, and are crucial in the formation of the chemical elements. In addition, much of what we know about neutrino oscillations is established from neutrino emerging from the Sun following the $\beta$-decay of reactions products, particularly $^{8}$B. The light-ion fusion reactions which encompass the Standard solar model need to be understood better, if solar neutrinos are to provide even more precise information on the neutrino oscillation properties. As an example,  the $^{7}$Be$(p,\gamma)^{8}$B radiative capture is a rather poorly-known step in the nucleosynthetic chain leading to $^{8}$B, which in turn is the dominant source of the high-energy solar neutrinos (through $\beta$-decay to $^8$Be) detected in terrestrial experiments.

Furthermore, nuclear reactions are one of the best tools for studying exotic nuclei, which have become the focus of the next generation experiments with rare-isotope beams. 
These are nuclei for which most low-lying states are unbound, so that a rigorous analysis requires scattering conditions. In addition, much of the information we have on the structure of such short-lived nuclei is inferred from reactions with other nuclei. 

Unfortunately, the calculation of nuclear reactions represent also a formidable challenge for nuclear theory, the main obstacle being the treatment of the scattering states. In this paper we will present a brief overview of existing theoretical methods for nuclear reactions, highlighting in particular recent progress in the {\em ab initio} calculation of low-energy scattering of light nuclei.

\subsection{ Overview of reaction approaches }
\label{QuaglioniS_Overview}
Because of their importance nuclear reactions attract much attention, and there have been many interesting new developments in the recent past. In this section we will give a brief overview of the theoretical efforts devoted to nuclear reactions, and in particular scattering of light nuclei. However, this is not intended to be completely exhaustive. 

Nuclear reaction approaches may be classified according to two broad categories. The first category embraces the so-called microscopic approaches, in which all the nucleons involved in the scattering process are active degrees of freedom, and the antisymmetrization of the many-body wave functions is treated exactly. 

In the three- and four-nucleon sectors there has been remarkable progress in the past ten years: the Faddeev~\cite{QuaglioniS_Witala01}, Faddeev-Yacubovski~\cite{QuaglioniS_Lazauskas05-1,QuaglioniS_Lazauskas09}, Alt-Grassberger and Sandhas \\(AGS)~\cite{QuaglioniS_Deltuva07-1,QuaglioniS_Deltuva07-2}, hyperspherical harmonics~\cite{QuaglioniS_Viviani09}, Lorentz integral transform methods~\cite{QuaglioniS_Gazit,QuaglioniS_LITNCSM,QuaglioniS_Bacca09}, etc., are among the best known of several numerically exact techniques able to describe reactions observables starting from realistic nucleon-nucleon ($NN$) and in some cases also three-nucleon ($NNN$) forces.

Going beyond four nucleons there are fewer {\em ab initio} or {\em ab initio} inspired methods able to describe reactions observables starting from realistic forces. Only very recently the Green's function Monte Carlo~\cite{QuaglioniS_GFMC_nHe4}, the no-core-shell model combined with the resonating group method (NCSM/RGM)~\cite{QuaglioniS_NCSMRGM_letter,QuaglioniS_NCSMRGM_prc} and the fermionic molecular dynamics~\cite{QuaglioniS_FMD} have made steps in this direction.

Reactions among light nuclei are more widely described starting from semirealistic $NN$ interactions with adjusted parameters within the traditional resonating-group method \cite{QuaglioniS_RGM,QuaglioniS_RGM1,QuaglioniS_RGM2,QuaglioniS_RGM3,QuaglioniS_Lovas98} or the generator coordinate method~\cite{QuaglioniS_Descouvemont,QuaglioniS_GCM,QuaglioniS_GCM3bcont}, which are microscopic cluster techniques.

A second category is that of few-body methods describing scattering among  structureless clusters. Here one starts form nucleon-Nucleus (usually optical) potentials fitted on some reaction observable, and the nucleus core is usually inert. There are exact techniques which can employ either local or non-local optical potentials like the Faddeev or AGS methods adopted by Deltuva~\cite{QuaglioniS_Deltuva09-1,QuaglioniS_Deltuva09-2}, and various approximated ones, like the continuum-discretized coupled channel equations~\cite{QuaglioniS_CDCC,QuaglioniS_XCDCC}, distorted wave born approximations, or various adiabatic approximations~\cite{QuaglioniS_adiabatic}, etc., which usually adopt local optical potentials.

Finally, there are also some recent attempts to describe reactions among light nuclei in an effective-field theory approach for halo nuclei~\cite{QuaglioniS_EFT-1,QuaglioniS_EFT-2}. Starting from experimental resonance parameters for the system under investigation, phase shifts and cross section are predicted at low energy.
 
Here we focus on one of the above mentioned approaches, {\em i.e.} the NSCM/RGM, for which we will present formalism in Sec.~\ref{QuaglioniS_formalism}, and a collection of results in Sec.~\ref{QuaglioniS_results}. Conclusions and an outline of possible future developments will be given in Sec.~\ref{QuaglioniS_conclusions}.

\section{Formalism}
\label{QuaglioniS_formalism}

A brief overview of the NCSM approach is presented in Sec.~\ref{QuaglioniS_ncsm}, the resonating-group method  is introduced in Sec.~\ref{QuaglioniS_rgm}, and the NCSM/RGM formalism is described in Section~\ref{QuaglioniS_ncsmrgm}.

\subsection{Ab initio no-core shell model}
\label{QuaglioniS_ncsm}
The NCSM is a technique for the solution of the $A$-nucleon bound-state problem. All $A$ (point-like) nucleons are active degrees of freedom, hence the difference with respect to standard shell model calculations with inert core. Starting from a microscopic Hamiltonian ($\vec{p}_i$ being the momentum of the ith nucleon and $m$ the nucleon mass)
\begin{equation}\label{ham}
H = 
\frac{1}{A}\sum_{i<j}\frac{(\vec{p}_i-\vec{p}_j)^2}{2m}
+ \sum_{i<j}^A V^{NN}_{ ij} + \sum_{i<j<k}^A V^{NNN}_{ijk} \; ,
\end{equation}
containing realistic $NN$ ($V^{NN}_{ij})$ or  $NN$ plus $NNN$ ($V^{NNN}_{ijk}$) forces (both coordinate- and momentum-space interactions can be equally handled), 
the non-relativistic Schr\"odinger equation is solved by expanding the wave functions in terms of a complete set of $A$-nucleon harmonic oscillator (HO) basis states up to a maximum excitation $N_{max}\hbar\Omega$ above the minimum energy configuration, with $\Omega$ the HO frequency.  

The choice of the HO basis in a complete $N_{max}\hbar\Omega$ space is motivated by its versatility. Indeed, this is the only basis
which allows to work within either Jacobi relative coordinates or Cartesian single-particle coordinates (as well as easily switch between the two), while preserving the translational invariance of the system. Consequently, powerful techniques based on the second quantization and developed 
for standard shell model calculations can be utilized.  As a downside, one has to face the consequences of the incorrect asymptotic behavior of the HO basis. 

Standard, accurate $NN$ potentials, such as the Argonne V18 (AV18)~\cite{QuaglioniS_AV18}, CD-Bonn~\cite{QuaglioniS_cdb2k}, INOY (inside non-local outside Yukawa)~\cite{QuaglioniS_INOY} and, to some extent, also the chiral N$^3$LO~\cite{QuaglioniS_N3LO}, generate strong short-range correlations that cannot be accommodated even in a reasonably large HO basis.  
In order to account for these short-range correlations and to accelerate convergence with respect to the increasing model space, the NCSM makes use of an effective interaction
obtained from the original, realistic $NN$ or $NN+NNN$ potentials by means of a unitary transformation in a $n-$body cluster approximation, where $n$ is typically $2$ or $3$~\cite{QuaglioniS_NO0203}.
The effective interaction depends on the basis truncation and by construction converges to
the original realistic $NN$ or $NN+NNN$ interaction as the size of the basis approaches infinity.

On the other hand, a new class of soft potentials has been recently developed, mostly by means of unitary transformations of the standard accurate $NN$ potentials mentioned above. These include the $V_{low {\it k}}$~\cite{QuaglioniS_Vlowk}, the Similarity Renormalization Group (SRG)~\cite{QuaglioniS_SRG} and the UCOM~\cite{QuaglioniS_UCOM} $NN$ potentials. A different class of soft phenomenological $NN$ potential used in some NCSM calculations are the so-called JISP potentials~\cite{QuaglioniS_JISP}. 
These so-called soft potentials are to some extent already renormalized for the purpose of simplifying many-body calculations. Therefore, one can perform convergent NCSM calculations with these potentials unmodified, or ``bare.'' In fact, the chiral N$^3$LO $NN$ potential~\cite{QuaglioniS_N3LO} can also be used bare with some success. NCSM calculations with bare potentials are variational with respect to the HO frequency and the basis truncation parameter $N_{\rm max}$.

\subsection{Resonating-group method}
\label{QuaglioniS_rgm}
The resonating-group method (RGM)~\cite{QuaglioniS_RGM,QuaglioniS_RGM1,QuaglioniS_RGM2,QuaglioniS_RGM3,QuaglioniS_Lovas98,QuaglioniS_Hofmann08} is a microscopic cluster technique in which the many-body Hilbert space is spanned by wave functions describing a system of two or more clusters in relative motion. Here, we will limit our discussion to the two-cluster RGM, which is based on binary-cluster channel states of total angular momentum $J$, parity $\pi$, and isospin $T$,
\begin{eqnarray}
|\Phi^{J^\pi T}_{\nu r}\rangle &=& \Big [ \big ( \left|A{-}a\, \alpha_1 I_1^{\,\pi_1} T_1\right\rangle \left |a\,\alpha_2 I_2^{\,\pi_2} T_2\right\rangle\big ) ^{(s T)}\nonumber\\
&&\times\,Y_{\ell}\left(\hat r_{A-a,a}\right)\Big ]^{(J^\pi T)}\,\frac{\delta(r-r_{A-a,a})}{rr_{A-a,a}}\,.\label{basis}
\end{eqnarray}
In the above expression, $\big ( \left|A{-}a\, \alpha_1 I_1^{\,\pi_1} T_1\right\rangle$ and $\left |a\,\alpha_2 I_2^{\,\pi_2} T_2\right\rangle$ are the internal (antisymmetric) wave functions of the first and second clusters, containing $A{-}a$ and $a$ nucleons ($a{<}A$), respectively. They are characterized by angular momentum quantum numbers $I_1$ and $I_2$ coupled together to form channel spin $s$. For their parity, isospin and additional quantum numbers we use, respectively, the notations $\pi_i, T_i$, and $\alpha_i$, with $i=1,2$. The cluster centers of mass are separated by the relative coordinate 
\begin{equation}
\vec r_{A-a,a} = r_{A-a,a}\hat r_{A-a,a}= \frac{1}{A - a}\sum_{i = 1}^{A - a} \vec r_i - \frac{1}{a}\sum_{j = A - a + 1}^{A} \vec r_j\,,
\end{equation}
where $\{\vec{r}_i, i=1,2,\cdots,A\}$ are the $A$ single-particle coordinates.
The channel states~(\ref{basis}) have relative angular momentum $\ell$. It is convenient to group all relevant quantum numbers into a cumulative index $\nu=\{A{-}a\,\alpha_1I_1^{\,\pi_1} T_1;\, a\, \alpha_2 I_2^{\,\pi_2} T_2;$ $\, s\ell\}$.
 
The former basis states can be used to expand the many-body wave function according to
\begin{equation}
|\Psi^{J^\pi T}\rangle = \sum_{\nu} \int dr \,r^2\frac{g^{J^\pi T}_\nu(r)}{r}\,\hat{\mathcal A}_{\nu}\,|\Phi^{J^\pi T}_{\nu r}\rangle\,. \label{trial}
\end{equation}
However, to preserve the Pauli principle one has to introduce the appropriate inter-cluster antisymmetrizer, schematically
\begin{equation}
\hat{\mathcal A}_{\nu}=\sqrt{\frac{(A{-}a)!a!}{A!}}\sum_{P}(-)^pP\,,
\end{equation}   
where the sum runs over all possible permutations $P$ 
that can be carried out 
among nucleons pertaining to different clusters, and $p$ is the number of interchanges characterizing them. 
Indeed, the basis states~(\ref{basis}) are not anti-symmetric under exchange of nucleons belonging to different clusters.

The coefficients of the expansion~(\ref{trial}) are the relative-motion wave functions $g^{J^\pi T}_\nu(r)$. These are the only unknowns of the problem, to be determined  solving the non-local integral-differential coupled-channel equations 
\begin{equation}
\sum_{\nu}\int dr \,r^2\left[{\mathcal H}^{J^\pi T}_{\nu^\prime\nu}(r^\prime, r)-E\,{\mathcal N}^{J^\pi T}_{\nu^\prime\nu}(r^\prime,r)\right] \frac{g^{J^\pi T}_\nu(r)}{r} = 0\,,\label{RGMeq}
\end{equation}
where $E$ denotes the total energy in the center-of-mass frame.
Here, the two integration kernels, specifically the Hamiltonian kernel,
\begin{equation}
{\mathcal H}^{J^\pi T}_{\nu^\prime\nu}(r^\prime, r) = \left\langle\Phi^{J^\pi T}_{\nu^\prime r^\prime}\right|\hat{\mathcal A}_{\nu^\prime}H\hat{\mathcal A}_{\nu}\left|\Phi^{J^\pi T}_{\nu r}\right\rangle\,,\label{H-kernel}
\end {equation}
and the norm kernel,
\begin{equation}
{\mathcal N}^{J^\pi T}_{\nu^\prime\nu}(r^\prime, r) = \left\langle\Phi^{J^\pi T}_{\nu^\prime r^\prime}\right|\hat{\mathcal A}_{\nu^\prime}\hat{\mathcal A}_{\nu}\left|\Phi^{J^\pi T}_{\nu r}\right\rangle\,,\label{N-kernel}
\end{equation}
contain all the nuclear structure and antisymmetrization properties of the problem. The somewhat unusual presence of a norm kernel is the result of the non-orthogonality of the basis states~(\ref{basis}), caused by the presence of the inter-cluster antisymmetrizer. The exchange terms of this antisymmetrization operator are also responsible for  the non-locality of the two kernels.

The main inputs of the RGM method are $i)$ the internucleon interaction;  and $ii)$ the wave functions of the $(A-a)$- and $a$-nucleon clusters.
Staring from the latters, in the traditional RGM the clusters internal wave functions are often (but not exclusively) translationally invariant HO shell-model functions of the lowest configuration or a linear superposition of such functions. The value of the HO size parameters $b$ are chosen ad hoc to reproduce properties of the nucleon clusters (such as size and/or binding energy, etc.) within the adopted interaction. This somewhat simplified description of the clusters internal wave functions is usually compensated by the use of semirealistic $NN$ interactions, such as the Volkov~\cite{QuaglioniS_Volkov} or Minnesota~\cite{QuaglioniS_Minnesota} potentials, with parameters that can be adjusted to reproduces important properties of the compound nucleus or reaction under study. The spin-orbit force, not present in the mentioned semi-realistic interactions, is sometimes added to the microscopic Hamiltonian. Exception to this general description of the RGM approach exist, particularly in the few-nucleon sector, where the method has been utilized in combination with realistic $NN$ and $NN+NNN$ forces~\cite{QuaglioniS_Hofmann08}. Finally, the treatment of the Coulomb interaction between charged clusters does not represent an issue in the RGM approach. 

The advantage in expressing the RGM basis states~(\ref{basis}) as antisymmetrized products of single-particle functions, and in particular Slater determinants, lies in the ability to carry out analytical derivations of the required matrix elements~(\ref{H-kernel}) and (\ref{N-kernel}).  Once the integration kernels are calculated, by solving the integral-differential coupled channel equations~(\ref{RGMeq}) subject to appropriate boundary conditions, one obtains bound-state wave functions and binding energies or scattering wave functions and  scattering matrix, from which any other scattering and reaction observable can be calculated.

\subsection{{\em Ab initio} NCSM/RGM approach}

\label{QuaglioniS_ncsmrgm}
A new first-principles, many-body approach capable of simultaneously describing both bound and scattering states in light nuclei has been developed by combining the RGM with the {\em ab initio} NCSM~\cite{QuaglioniS_NCSMRGM_letter,QuaglioniS_NCSMRGM_prc}. This new approach complements the microscopic-cluster technique of the RGM with the utilization of realistic interactions and a consistent microscopic description of the nucleonic clusters, while preserving important symmetries such as Pauli exclusion principle, translational invariance, and angular momentum.
More in detail, the formalism presented in Section~\ref{QuaglioniS_rgm} can be combined with the {\em ab initio} NCSM as follows. 

First, we note that the Hamiltonian can be written as
\begin{equation}\label{Hamiltonian}
H=T_{\rm rel}(r)+ {\mathcal V}_{\rm rel} +\bar{V}_{\rm C}(r)+H_{(A-a)}+H_{(a)}\,,
\end{equation}
where $H_{(A-a)}$ and $H_{(a)}$ are the ($A{-}a$)- and $a$-nucleon intrinsic Hamiltonians, respectively, $T_{\rm rel}(r)$ is the relative kinetic energy 
and ${\mathcal V}_{\rm rel}$ is the sum of all interactions between nucleons belonging to different clusters after subtraction of the average Coulomb interaction between them, explicitly singled out in the term $\bar{V}_{\rm C}(r)=Z_{1\nu}Z_{2\nu}e^2/r$ ($Z_{1\nu}$ and $Z_{2\nu}$ being the charge numbers of the clusters in channel $\nu$)
\begin{eqnarray}
{\mathcal V}_{\rm rel} &=& \sum_{i=1}^{A-a}\sum_{j=A-a+1}^AV^{NN}_{ij}
-\bar{V}_{\rm C}(r)
+{\mathcal V}^{NNN}_{(A-a,a)}
\nonumber\\[2mm]
&=& \sum_{i=1}^{A-a}\sum_{j=A-a+1}^A \Big[V^{N}(\vec r_i-\vec r_j, \sigma_i,\sigma_j,\tau_i,\tau_j)\nonumber\\[2mm]
&&  + \frac{e^2(1+\tau^z_i)(1+\tau^z_j)}{4|\vec r_i-\vec r_j|} -\frac{1}{(A-a)a}\bar V_{\rm C}(r)\Big]\nonumber\\[2mm]
&&+ {\mathcal V}^{NNN}_{(A-a,a)}\label{pot}\,.
\end{eqnarray}
Nuclear, $V^{N}(\vec r_i-\vec r_j, \sigma_i,\sigma_j,\tau_i,\tau_j)$, and point-Coulomb components of the two-body potential have been listed explicitly ( $\sigma_i$, $\tau_i$ denoting spin and isospin coordinates, respectively, of the ith nucleon). If the $A$-nucleon Hamiltonian contains a $NNN$ force, the inter-cluster interaction ${\mathcal V}_{\rm rel}$ will present also a contribution from the latter, denoted here with $V^{NNN}_{(A-a,a)}$.

The cluster's Hamiltonians and inter-cluster interaction ${\mathcal V}_{\rm rel}$ are consistent, as they contain the same realistic potentials. The clusters internal wave functions are also treated consistently: $\big ( \left|A{-}a\, \alpha_1 I_1^{\,\pi_1} T_1\right\rangle$ and $\left |a\,\alpha_2 I_2^{\,\pi_2} T_2\right\rangle$ are obtained by diagonalizing $H_{(A-a)}$ and $H_{(a)}$, respectively, in the model spaces spanned by the NCSM basis. Both the $(A-a)$- and $a$-nucleon model spaces are characterized by the same HO frequency $\Omega$ and maximum number $N_{\rm max}$ of excitations above the minimum configuration.  At the same time,  in calculating the Hamiltonian and norm kernels of Eqs.~(\ref{H-kernel}), and~(\ref{N-kernel}), all ``direct'' terms arising from the identical permutations in both $\hat{\mathcal A}_{\nu}$ and $\hat{\mathcal A}_{\nu^\prime}$ are treated exactly (with respect to the separation $r$) 
with the exception of  $\left\langle\Phi^{J^\pi T}_{\nu^\prime r^\prime}\right|{\mathcal V}_{\rm rel}\left|\Phi^{J^\pi T}_{\nu r}\right\rangle$. The latter and all remaining terms are localized and can be  obtained by expanding the Dirac $\delta$ of Eq.~(\ref{basis}) on a set of HO radial wave functions with identical frequency $\Omega$, and model-space size $N_{\rm max}$ consistent with those used for the two clusters. In this respect, we note that,  thanks to the subtraction of the average potential $\bar{V}_{\rm C}(r)$, ${\mathcal V}_{\rm rel}$  is localized also in the presence of the Coulomb force.

If the adopted potential generates strong short-range correlations, we employ consistent NCSM effective interactions derived from it. More specifically, the cluster eigenstates are obtained by employing the usual NCSM effective interaction~\cite{QuaglioniS_NO0203}. However, in place of the bare potential entering ${\mathcal V}_{\rm rel}$ we adopt a modified effective interaction, which avoids renormalizations related to the kinetic energy.  Following the notation of Ref.~\cite{QuaglioniS_NO0203}, at the two-body cluster level this is given by $V^\prime_{2\rm{eff}}=\bar{H}_{2\rm{eff}}-\bar{H}^\prime_{2\rm{eff}}$, where $\bar{H}^\prime_{2\rm{eff}}$ is the effective Hamiltonian derived from $H^{\Omega\,\prime}_2 = H_{02}+V^\prime_{12}$, with $V^\prime_{12} = - m\Omega^2\vec{r}^{\,2}/A$. 
Note that $V^\prime_{2\rm{eff}}\rightarrow V_{N}$ in the limit $N_{\rm max}\rightarrow\infty$ and, for each model space, the renormalizations related to the kinetic energy and the HO potential introduced in  $\bar H_{2\rm eff}$ are compensated by the subtraction of $\bar H^\prime_{2\rm eff}$. The kinetic-energy renormalizations are appropriate within the standard NCSM, but they would compromise scattering results obtained within the NCSM/RGM approach, in which the relative kinetic energy and average Coulomb potential between the clusters are treated exactly (that is, are not truncated within a finite HO model space).

\subsubsection{Integration kernels}
\label{QuaglioniS_kernels}

\begin{figure*}[!htb]
\begin{minipage}{6.5cm}
\rotatebox{-90}{ \includegraphics[height=.24\textheight]{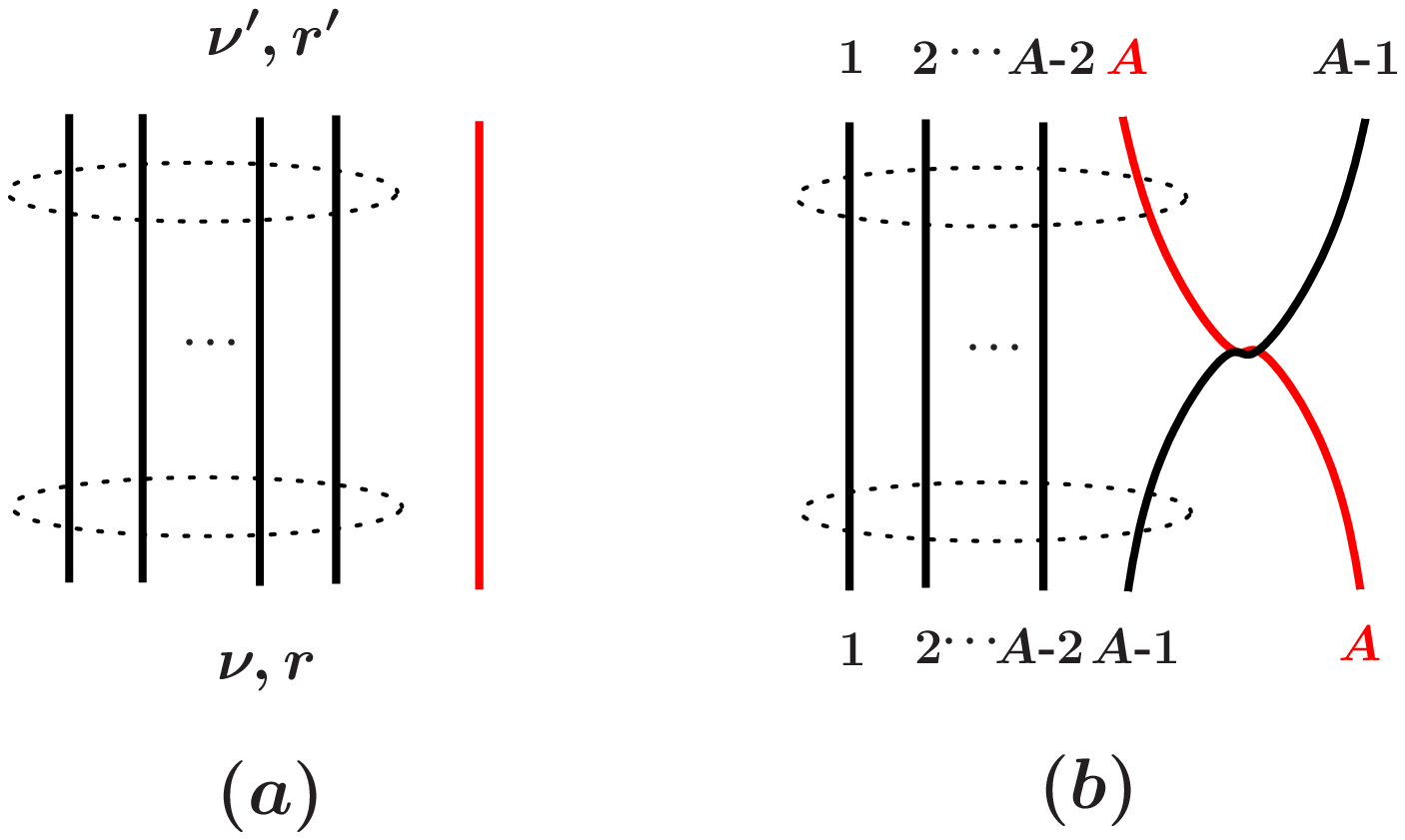}}
\end{minipage}
\begin{minipage}{9.5cm}
\rotatebox{-90}{\includegraphics[height=.39\textheight]{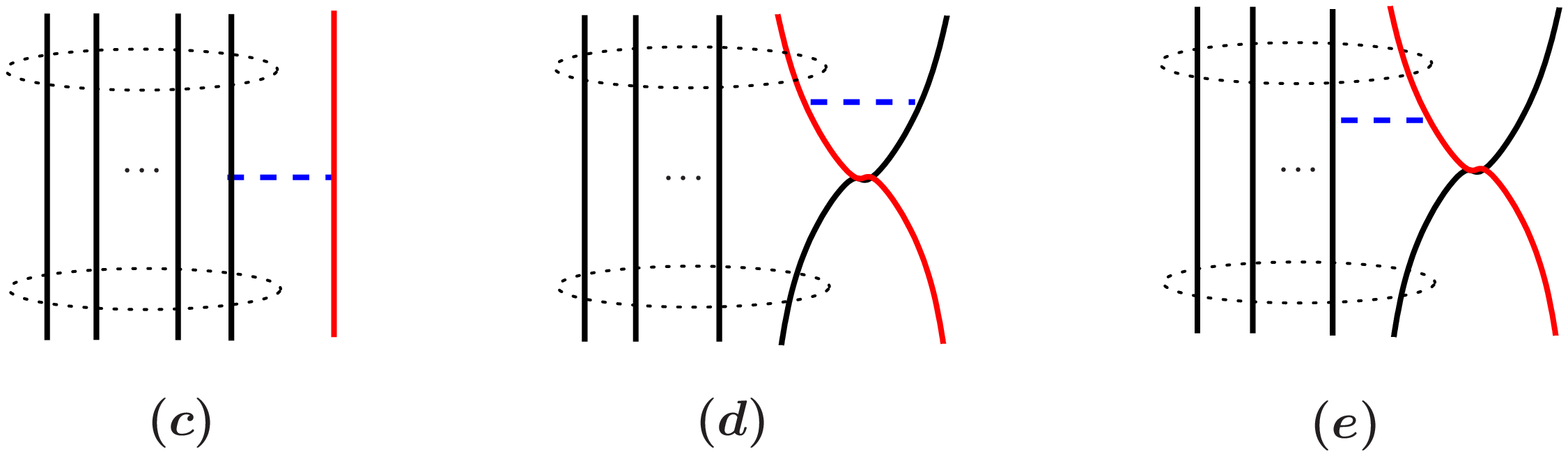}}
\end{minipage}
\caption{Diagrammatic representation of: ($a$) ``direct" and ($b$) ``exchange"  components of the norm kernel; ($c$ and $d$) ``direct"  and ($e$) ``exchange"   components of the potential kernel for the $(A-1,1)$ cluster basis. The first group of circled lines represents the first cluster, the bound state of $A{-}1$ nucleons. The separate line represents the second cluster, in the specific case  a single nucleon. Bottom and upper part of the diagram represent initial and final states, respectively.}\label{QuaglioniS_fig:1}
\end{figure*}

To give a somewhat more in depth description of the formalism involved in the calculation of the matrix elements (\ref{H-kernel}) and (\ref{N-kernel}), here we will present examples of algebraic expressions derived within the single-nucleon projectile basis, i.e., for binary-cluster channel states~(\ref{basis}) with $a=1$ (with channel index $\nu = \{ A{-}1 \, \alpha_1 I_1^{\pi_1} T_1; \, 1\, \frac 1 2 \frac 1 2;\, s\ell\}$). In this model space, the norm kernel is rather simple and is given by 
\begin{eqnarray}
{\mathcal N}^{J^\pi T}_{\nu^\prime\nu}(r^\prime, r)& = &\left\langle\Phi^{J^\pi T}_{\nu^\prime r^\prime}\right|1-\sum_{i=1}^{A-1}\hat P_{iA} \left|\Phi^{J^\pi T}_{\nu r}\right\rangle
\\
&=&\delta_{\nu^\prime\,\nu}\,\frac{\delta(r^\prime-r)}{r^\prime\,r}-(A-1)\sum_{n^\prime n}R_{n^\prime\ell^\prime}(r^\prime) R_{n\ell}(r)\nonumber\\
&&\times \left\langle\Phi^{J^\pi T}_{\nu^\prime n^\prime}\right|\hat P_{A-1,A} \left|\Phi^{J^\pi T}_{\nu n}\right\rangle\,,\label{norm}
\end{eqnarray}
where it is easy to recognize a direct term, in which initial and final state are identical (corresponding to diagram $(a)$ of Fig.~\ref{QuaglioniS_fig:1}), and a many-body correction due to the exchange part of the inter-cluster anti-symmetrizer (corresponding to diagram $(b)$ of Fig.~\ref{QuaglioniS_fig:1}). 
As the exchange $\hat P_{A-1,A}$  is a short-range operator, in calculating its matrix elements  we replaced the delta function of Eq.~(\ref{basis}) with its representation in the HO model space. Such HO expansion is appropriate whenever the operator is short-to-medium range.

The presence of the inter-cluster anti-symmetrizer affects also the Hamiltonian kernel, and in particular the matrix elements of the interaction:
\begin{eqnarray}
{\mathcal H}^{J^\pi T}_{\nu^\prime\nu}(r^\prime, r) &\!=\!& \left\langle\Phi^{J^\pi T}_{\nu^\prime r^\prime}\right|H\Big[1-\sum_{i=1}^{A-1}\hat P_{iA}\Big] \left|\Phi^{J^\pi T}_{\nu r}\right\rangle\label{ham}\\
&\!=\!& \Big[\hat T_{\rm rel}(r^\prime)\!+\!\bar V_{\rm C}(r^\prime)\!+\!E^{I_1^{\prime\pi^\prime_1}T^\prime_1}_{\alpha^\prime_1}\Big]\,{\mathcal N}^{J^\pi T}_{\nu^\prime\nu}(r^\prime, r)\nonumber\\
&&\nonumber\\
&&+{\mathcal V}^{\rm D}_{\nu^\prime\nu}(r^\prime,r)+{\mathcal V}^{\rm \,ex}_{\nu^\prime\nu}(r^\prime,r)\,,
\end{eqnarray}
where $E^{I_1^{\prime\pi^\prime_1}T^\prime_1}_{\alpha^\prime_1}$ are the eigenenergies of the $(A-1)$-nucleon cluster. If no $NNN$ forces are present in the Hamiltonian one obtains a ``direct" term involving interaction and exchange of one of the nucleons in the first cluster with the nucleon ($a=1$) of the second cluster (see diagrams ($c$) and ($d$) of Fig.~\ref{QuaglioniS_fig:1}), and an ``exchange" term involving the interaction of the $A$th nucleon with one of the $(A-1)$ nucleons, accompanied by the exchange with a second of such nucleons. Diagram ($e$) of Fig.~\ref{QuaglioniS_fig:1} describes this latter term. 
These two potential kernels, which together constitute the matrix element $\left\langle\Phi^{J^\pi T}_{\nu^\prime r^\prime}\right|{\mathcal V}_{\rm rel}\,\hat{\mathcal A}^2\left|\Phi^{J^\pi T}_{\nu r}\right\rangle$, have the following expressions:
\begin{eqnarray}
{\mathcal V}^{\rm D}_{\nu^\prime\nu}(r^\prime,r) &=&(A-1)\sum_{n^\prime n}R_{n^\prime\ell^\prime}(r^\prime) R_{n\ell}(r)\nonumber\\
&&\times \left\langle\Phi^{J^\pi T}_{\nu^\prime n^\prime}\right|V_{A-1,A}\big(1\!-\!\hat P_{A-1,A}\big) \left|\Phi^{J^\pi T}_{\nu n}\right\rangle\nonumber\\
&&\label{D-potential}\\
{\mathcal V}^{\rm\, ex}_{\nu^\prime\nu}(r^\prime,r) &=&-(A-1)(A-2)\sum_{n^\prime n}R_{n^\prime\ell^\prime}(r^\prime) R_{n\ell}(r)\nonumber\\
&&\times \left\langle\Phi^{J^\pi T}_{\nu^\prime n^\prime}\right|\hat P_{A-1,A}\,V_{A-2,A-1} \left|\Phi^{J^\pi T}_{\nu n}\right\rangle\!. \label{ex-potential}
\end{eqnarray} 
The inclusion of a $NNN$ interaction in the Hamiltonian is straightforward, and amounts to extra ``direct" and ``exchange" potential kernels, which can be obtained in a similar way. 

Being translationally-invariant  quantities, the Hamiltonian and norm kernels~(\ref{H-kernel},~\ref{N-kernel}) can be ``naturally" derived working within the NCSM Jacobi-coordinate basis. However, particularly for the purpose of calculating reactions involving $p$-shell nuclei, it is computationally advantageous to introduce  Slater-determinant (SD) channel states of the type 
\begin{eqnarray}
|\Phi^{J^\pi T}_{\nu n}\rangle_{\rm SD}   &=&    \Big [\big (\left|A{-}a\, \alpha_1 I_1 T_1\right\rangle_{\rm SD} 
\left |a\,\alpha_2 I_2 T_2\right\rangle\big )^{(s T)}\nonumber\\
&&\times Y_{\ell}(\hat R^{(a)}_{\rm c.m.})\Big ]^{(J^\pi T)} R_{n\ell}(R^{(a)}_{\rm c.m.})\,,
\label{SD-basis}
\end{eqnarray}
in which the eigenstates of the $(A{-}a)$-nucleon fragment are obtained in the SD basis (while the second cluster is still a NCSM Jacobi-coordinate eigenstate), and $\vec{R}^{(a)} = a^{-1/2}\sum_{i = A-a+1}^A \vec{r}_i$ is the vector proportional to the center of mass coordinate of the $a$-nucleon cluster. Indeed, it easy to demonstrate that translationally invariant matrix elements can be extracted from those calculated in the SD basis of Eq.~(\ref{SD-basis}) by inverting the following expression:
 \begin{eqnarray}
&& {}_{\rm SD}\!\left\langle\Phi^{J^\pi T}_{\nu^\prime n^\prime}\right|\hat{\mathcal O}_{\rm t.i.}\left|\Phi^{J^\pi T}_{\nu n}\right\rangle\!{}_{\rm SD} = \nonumber\\
&&\nonumber\\
&&\sum_{n^\prime_r \ell^\prime_r, n_r\ell_r, J_r}
 \left\langle\Phi^{J_r^{\pi_r} T}_{\nu^\prime_r n^\prime_r}\right|\hat{\mathcal O}_{\rm t.i.}\left|\Phi^{J_r^{\pi_r} T}_{\nu_r n_r}\right\rangle\nonumber\\
&&  \times \sum_{NL} \hat \ell \hat \ell^\prime \hat J_r^2 (-1)^{(s+\ell-s^\prime-\ell^\prime)}
  \left\{\begin{array}{ccc}
 s &\ell_r&  J_r\\
  L& J & \ell
 \end{array}\right\}
 \left\{\begin{array}{ccc}
 s^\prime &\ell^\prime_r&  J_r\\
  L& J & \ell^\prime
 \end{array}\right\}\nonumber\\
 &&\nonumber\\
&& \times\langle  n_r\ell_rNL\ell | 00n\ell\ell \rangle_{\frac{a}{A-a}} 
 \;\langle  n^\prime_r\ell^\prime_rNL\ell | 00n^\prime\ell^\prime\ell^\prime \rangle_{\frac{a}{A-a}} \,.\label{Oti}
 \end{eqnarray}
Here $\hat {\mathcal O}_{\rm t.i.}$ represents any scalar and parity-conserving and translationally-invariant operator ($\hat {\mathcal O}_{\rm t.i.} = \hat{\mathcal A}$, $\hat{\mathcal A} H \hat{\mathcal A}$, etc.), and $\langle  n_r\ell_rNL\ell | 00n\ell\ell \rangle_{\frac{a}{A-a}}$, $\langle  n^\prime_r\ell^\prime_rNL\ell | 00n^\prime\ell^\prime\ell^\prime \rangle_{\frac{a}{A-a}}$ are general HO brackets for two particles with mass ratio $a/(A-a)$.
We exploited both Jacobi-coordinate and SD channel states to verify our results.  

As an example, the single-nucleon projectile ``exchange" part of the norm kernel within the Jacobi-coordinate basis for a system of $A=3$ nucleons is given by: 
\begin{eqnarray}
{\mathcal N}^{J^\pi T}_{\nu^\prime\nu}(r^\prime,r) &=&\delta_{\nu^\prime\nu}\frac{\delta(r^\prime-r)}{r^\prime\,r} -2\sum_{n^\prime n}R_{n^\prime\ell^\prime}(r^\prime)R_{n\ell}(r)\\\nonumber
&&\times\sum_{n^\prime_1\ell^\prime_1 s^\prime_1} \big\langle n^\prime_1\ell^\prime_1 s^\prime_1 I^\prime_1 T^\prime_1\big|2\alpha^\prime_1 
I_1^{\prime\pi^\prime_1} T^\prime_1\big\rangle \\\nonumber
&&\times\sum_{n_1\ell_1 s_1} \big\langle n_1\ell_1 s_1 I_1 T_1\big|2\alpha_1 I_1^{\pi_1} T_1\big\rangle\nonumber\\[2mm]
&& \times\hat T^\prime_1\hat T_1 (-)^{T^\prime_1+T_1}
\left\{\begin{array}{ccc}
s&\frac12&T_1\\[2mm]
\frac12&T&T^\prime_1
\end{array}\right\} 
\hat s^\prime_1\hat s_1\hat I_1^\prime\hat I_1\hat s^\prime\hat s \,(-)^{\ell_1 + \ell} \nonumber\\
&&\times \sum_{\Lambda,Z}\hat\Lambda^2\hat Z^2 (-)^\Lambda
\left\{\begin{array}{ccc}
\frac12&\frac12&s_1\\[2mm]
\frac12&Z&s^\prime_1
\end{array}\right\} 
\left\{\begin{array}{ccc}
\ell^\prime_1&Z&s^\prime\\[2mm]
J&\ell^\prime&\Lambda
\end{array}\right\} \nonumber\\[2mm]
&&\times\left\{\begin{array}{ccc}
\ell^\prime_1&Z&s^\prime\\[2mm]
\frac12&I^\prime_1&s^\prime_1
\end{array}\right\}  
\left\{\begin{array}{ccc}
\ell_1&Z&s\\[2mm]
J&\ell&\Lambda
\end{array}\right\}
\left\{\begin{array}{ccc}
\ell_1&Z&s\\[2mm]
\frac12&I_1&s_1
\end{array}\right\} \nonumber\\
&&\times \langle n^\prime\ell^\prime,n^\prime_1\ell^\prime_1,\Lambda|n_1\ell_1,n\ell,\Lambda\rangle_3\,. 
\label{norm-3}
\end{eqnarray}
Here 
$\big\langle n_1\ell_1 s_1 I_1 T_1\big|2\alpha_1 I_1^{\pi_1} T_1\big\rangle$ and  $ \big\langle n^\prime_1\ell^\prime_1 s^\prime_1 I^\prime_1 T^\prime_1\big|2\alpha^\prime_1 I_1^{\prime\pi^\prime_1} T^\prime_1\big\rangle$ are the coefficients of the expansion of initial and final two-nucleon target wave functions, respectively,  with respect to the HO basis states depending on the Jacobi, spin, and isospin coordinates $\boldsymbol{\xi}_1 = (\vec{r}_1 - \vec{r}_2)/\sqrt2$, $\sigma_1,\sigma_2$, and $\tau_1,\tau_2$, respectively, 
\begin{equation}
\langle\vec{\xi}_1\sigma_1\sigma_2\tau_1\tau_2|n_1\ell_1s_1I_1T_1\rangle\,,\label{2-basis}
\end{equation} 
where $n_1,\ell_1$ are the HO quantum numbers corresponding to the harmonic oscillator associated with $\vec\xi_1$, while $s_1,I_1$, and $T_1$ are the  spin, total angular momentum, and isospin of the two-nucleon channel formed by nucleons 1 and 2, respectively. Note that the basis~(\ref{2-basis}) is anti-symmetric with respect to the exchange of the two nucleons, $(-)^{\ell_1+s_1+T_1}=-1$. Finally, $\langle n^\prime\ell^\prime,n^\prime_1\ell^\prime_1,\Lambda|n_1\ell_1,n\ell,\Lambda\rangle_3$ are the general HO brackets for two particles with mass ratio $3$. 

\begin{figure}[!htb]
\centering
\includegraphics[width=0.9\columnwidth,angle=0]{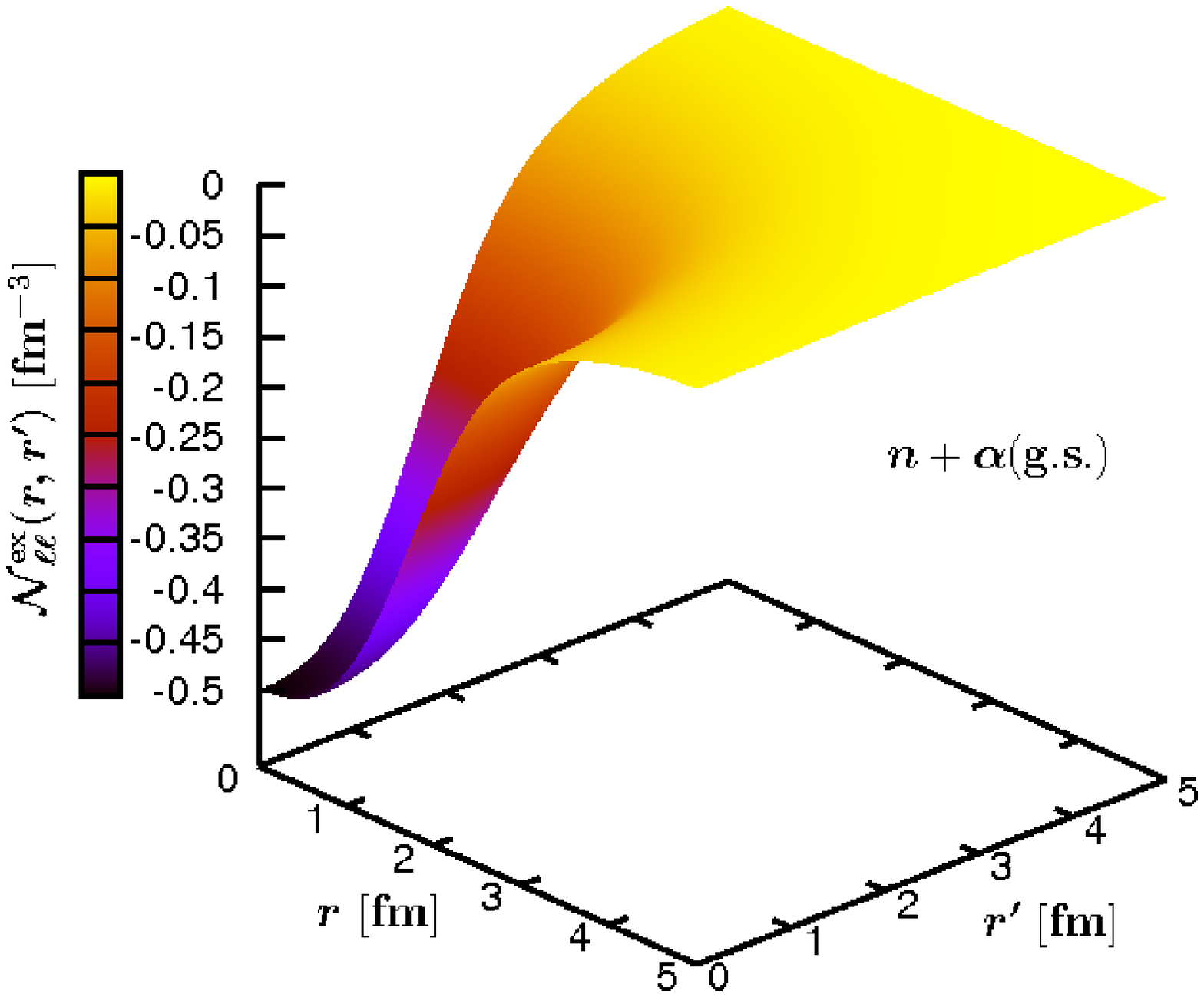}
\includegraphics[width=0.9\columnwidth,angle=0]{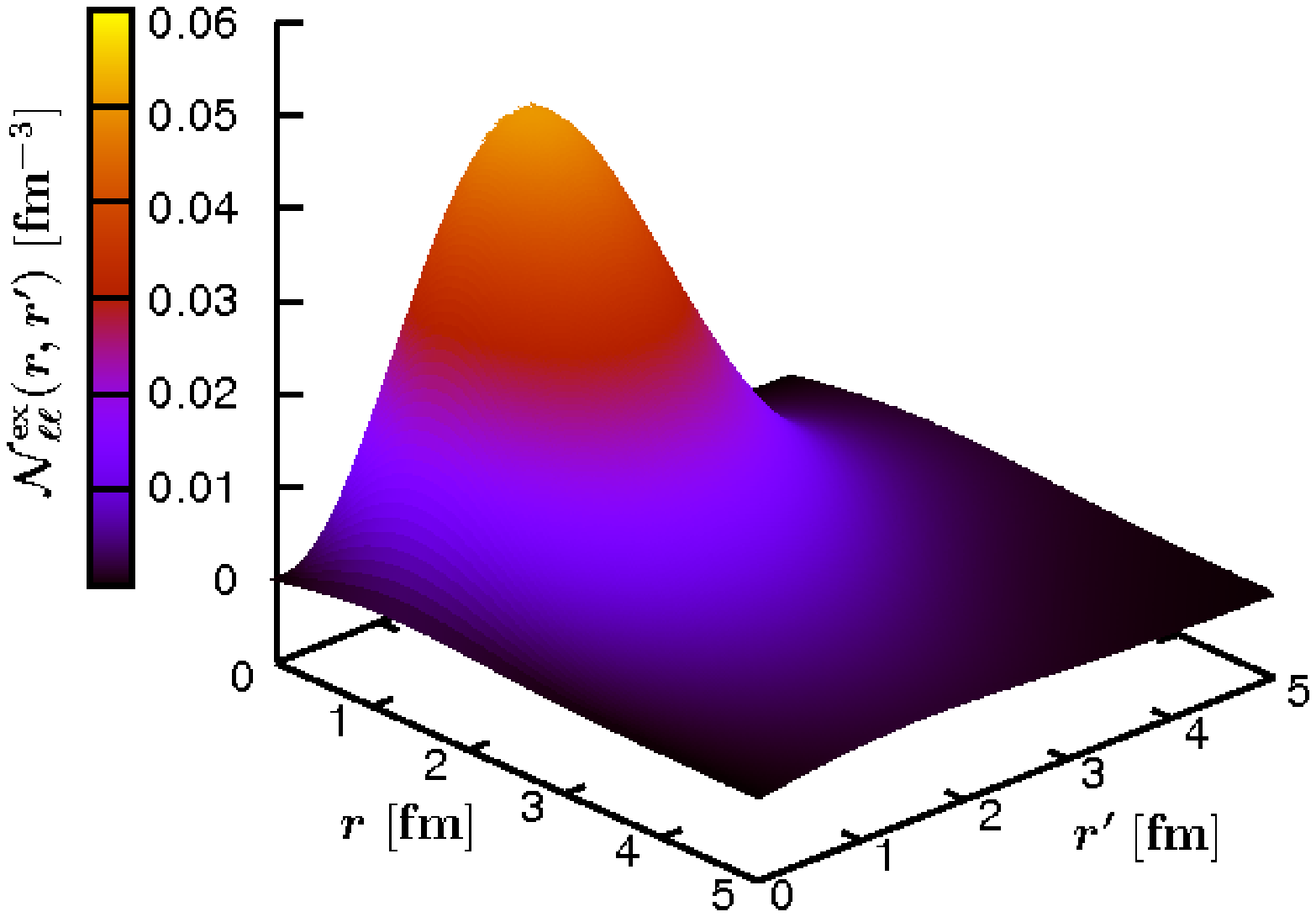}
\caption{Calculated ``exchange" part of the norm kernel, ${\mathcal N}^{J^\pi T}_{\ell\ell}(r^\prime,r) - \frac{\delta(r^\prime-r)}{r^\prime\,r}$ for the $n+\alpha$(g.s.) ${}^2S_{1/2}$ (top panel), and ${}^2P_{3/2}$ (bottom panel) channels as a function of the relative coordinates $r$ and $r^\prime$, using the N$^3$LO $NN$ potential~\cite{QuaglioniS_N3LO} at $\hbar\Omega=19$ MeV. The ${}^2S_{1/2}$ channel is strongly influenced by the Pauli-exclusion principle, which forbids to accommodate more than four nucleons into the $s$-shell of a nuclear system. The four nucleons forming the $^4$He g.s.  sit mostly in the $0\hbar\Omega$ shell. Accordingly, in the $^2S_{1/2}$ channel the ``exchange"-part of the norm kernel 
suppresses the (dominant) $0\hbar\Omega$ contribution to the $\delta$ function of Eq.~(\ref{norm}) (and, consequently, to the $S$-wave relative-motion wave function $g^{\frac12^+\frac12}_{\ell=0}$) coming from the fifth nucleon in $s$-shell configuration. 
The situation is different if we consider a different partial wave, like the ${}^2P_{3/2}$ channel, corresponding to the $^5$He resonance. In this case the exchange norm represents a small and positive correction to the delta function. 
}
\label{QuaglioniS_fig:2}       
\end{figure}
At the same time, as mentioned above, the matrix elements of the operators $\hat P_{A-1,A}$, $V_{A-1,A}(1-\hat P_{A-1,A})$, and $\hat P_{A-1,A}V_{A-2,A-1}$ can be more intuitively derived working\\ within the SD basis of Eq.~(\ref{SD-basis}). Using the second-quantization formalism, they can be related to linear combinations of matrix elements of creation and annihilation operators between $(A{-}1)$-nucleons SD states. As an example, the case of the exchange operator $\hat P_{A-1, A}$ yields:
\begin{eqnarray}
&&_{\rm SD}\langle \Phi_{\nu'\,n'}^{J^\pi T}|\hat P_{A,A-1}| \Phi_{\nu\,n}^{J^\pi T}\rangle_{\rm SD}\nonumber\\[2mm]
&&=\frac{1}{A-1} \sum_{jj'K\tau} 
\hat{s}\hat{s}'\hat{j}\hat{j}'\hat{K}\hat{\tau} (-1)^{I'_1+j'+J} (-1)^{T_1+\frac{1}{2}+T}\nonumber\\
&&\times\left\{ \begin{array}{@{\!~}c@{\!~}c@{\!~}c@{\!~}} 
I_1 & \frac{1}{2} & s \\[2mm] 
\ell & J & j 
\end{array}\right\} 
\left\{ \begin{array}{@{\!~}c@{\!~}c@{\!~}c@{\!~}} I'_1 & \frac{1}{2} & s' \\ [2mm]
\ell' & J & j' \end{array}\right\} \left\{ \begin{array}{@{\!~}c@{\!~}c@{\!~}c@{\!~}} 
I_1 & K & I'_1 \\[2mm] 
j' & J & j \end{array}\right\}
\left\{ \begin{array}{@{\!~}c@{\!~}c@{\!~}c@{\!~}} 
T_1 & \tau & T'_1 \\[2mm]
\frac{1}{2} & T & \frac{1}{2} \end{array}\right\}\nonumber\\[2mm]
&&\times\; _{\rm SD}\langle A{-}1 \alpha' I'_1 T'_1 ||| (a^\dagger_{n\ell j\frac{1}{2}} \tilde{a}_{n'\ell'j'\frac{1}{2}})^{(K\tau)} ||| A{-}1 \alpha I_1 T_1 \rangle_{\rm SD}\,.\nonumber\\
\label{P_AAm1_SD}
\end{eqnarray}
Here, 
$_{\rm SD}\langle A{-}1 \alpha' I'_1 T'_1 ||| (a^\dagger_{n\ell j\frac{1}{2}} \tilde{a}_{n'\ell'j'\frac{1}{2}})^{(K\tau)} ||| A{-}1 \alpha I_1 T_1 \rangle_{\rm SD}$ are one-body density matrix elements of the target nucleus
and 
$\tilde{a}_{n'\ell'j'm'\frac{1}{2}m_t^\prime}=(-1)^{j'-m'+\frac{1}{2}-m_t^\prime}\;a_{n'\ell'j'-m'\frac{1}{2}-m_t^\prime}$.
Next we extract  the corresponding translationally-invariant matrix elements, 
$\langle \Phi_{\nu^\prime_r\,n'_r}^{(A-1,1) J_r^{\pi_r}T}|\hat P_{A,A-1}| \Phi_{\nu_r\,n_r}^{(A-1,1) J_r^{\pi_r}T}\rangle$, 
by inverting Eq.~(\ref{Oti}) for $a=1$ and $\hat {\mathcal O}_{\rm t.i.}=\hat P_{A-1,A}$. 
The final step follows easily from Eq.~(\ref{norm}). 
\begin{figure*}[!htb]
\centering
\includegraphics[width=1.8\columnwidth,angle=0]{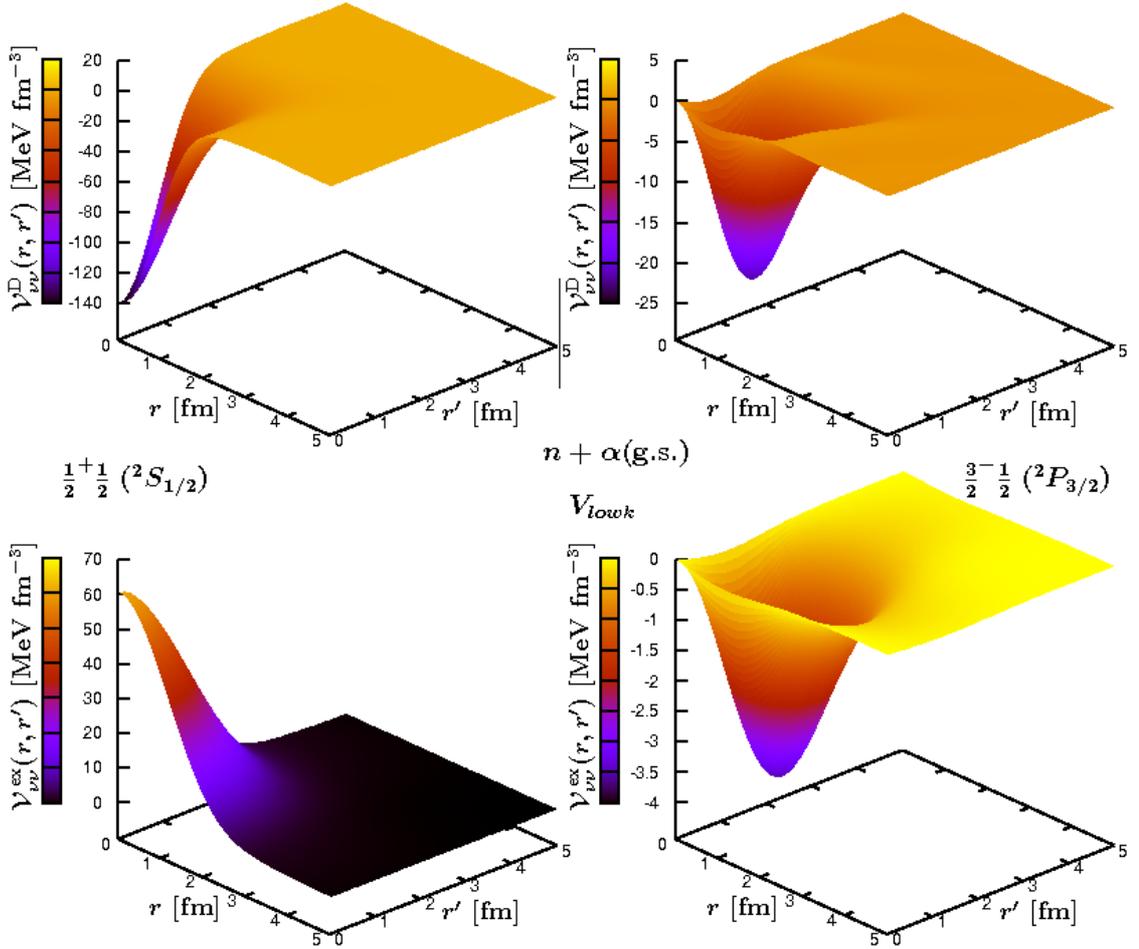}
\caption{Calculated ``direct" (top figures) and ``exchange" (bottom figures) potential kernels for the $n+\alpha$(g.s.) ${}^2S_{1/2}$ (left column), and ${}^2P_{3/2}$ (right column) channels as a function of the relative coordinates $r$ and $r^\prime$, using the $V_{{\rm low}k}$ $NN$ potential~\cite{QuaglioniS_Vlowk} at $\hbar\Omega=18$ MeV. The ${}^2S_{1/2}$ channel is strongly influenced by the Pauli-exclusion principle, which forbids to accommodate more than four nucleons into the $s$-shell of a nuclear system: the exchange potential introduces  repulsion between a nucleon in the $\alpha$ particle and the fifth nucleon when they are both in s-shell, largely suppressing the attractive contribution of the direct potential.
The situation is different if we consider a different partial wave, like the ${}^2P_{3/2}$ channel, corresponding to the $^5$He resonance. In this case 
both direct and exchange potential are attractive. 
}
\label{QuaglioniS_fig:3}       
\end{figure*}

Due to the exchange terms of the intercluster antisymmetrizers, norm and potential kernels are non-local and appear as surfaces in three dimensions such as, {\em e.g.}, those shown in Figs.~\ref{QuaglioniS_fig:2} and ~\ref{QuaglioniS_fig:3}. The latter figures present results of single-channel calculations carried out using $n$-$\alpha$ cluster channels with the $\alpha$ particle in its g.s. (note that the index $\nu=\{4\, {\rm g.s.}\, 0^+0; 1 \frac12^+\frac12; \frac12 \ell\}$ is simply replaced by the quantum number $\ell$). The interaction models adopted are the N$^3$LO $NN$ potential~\cite{QuaglioniS_N3LO}, and the $V_{{\rm low}k}$ $NN$ potential~\cite{QuaglioniS_Vlowk} derived from AV18~\cite{QuaglioniS_AV18} with cutoff $\Lambda=2.1$ fm$^{-1}$.
 
\subsubsection{Orthogonalization of the RGM equations}
\label{QuaglioniS_orthog}
An important point to notice, is that Eq.~(\ref{RGMeq}) does not represent a system of multichannel Schr\"odinger equations, and $g^{J^\pi T}_\nu(r)$ do not represent Schr\"odinger wave functions. This feature, which is indicated by the presence of the norm kernel ${\mathcal N}^{J^\pi T}_{\nu^\prime\nu}(r^\prime, r)$ and is caused by the short-range non-orthogonality induced by the non-identical permutations in the inter-cluster anti-symmetrizers,  can be removed by introducing normalized Schr\"odinger wave functions
\begin{equation}
\frac{\chi^{J^\pi T}_\nu(r)}{r} = \sum_{\gamma}\int dy\, y^2 {\mathcal N}^{\frac12}_{\nu\gamma}(r,y)\,\frac{g^{J^\pi T}_\gamma(y)}{y}\,,
\end{equation}
where ${\mathcal N}^{\frac12}$ is the square root of the norm kernel, and applying the inverse-square root of the norm kernel, ${\mathcal N}^{-\frac12}$, to both left and right-hand side of the square brackets in Eq.~(\ref{RGMeq}).  By means of this procedure, known as orthogonalization and explained in more detail in Ref.~\cite{QuaglioniS_NCSMRGM_prc}, one obtains a system of multichannel Schr\"odinger equations:
\begin{eqnarray}
&&[\hat T_{\rm rel}(r) + \bar V_{\rm C}(r) -(E - E_{\alpha_1}^{I_1^{\pi_1} T_1} - E_{\alpha_2}^{I_2^{\pi_2} T_2})]\frac{\chi^{J^\pi T}_{\nu} (r)}{r} \nonumber\\[2mm]
&&+ \sum_{\nu^\prime}\int dr^\prime\,r^{\prime\,2} \,W^{J^\pi T}_{\nu \nu^\prime}(r,r^\prime)\,\frac{\chi^{J^\pi T}_{\nu^\prime}(r^\prime)}{r^\prime} = 0,\label{r-matrix-eq}
\end{eqnarray} 
where $E_{\alpha_i}^{I_i^{\pi_i} T_i}$ are the energy eigenvalues of the $i$-th cluster ($i=1,2$), and $W^{J^\pi T}_{\nu^\prime \nu}(r^\prime,r)$ are the overall non-local potentials between the two clusters, which depend upon the channel of relative motion, while do not depend upon the energy $E$ of the system.     

\section{Results}
\label{QuaglioniS_results}
\begin{figure}[!htb]
\centering
\includegraphics[width=0.8\columnwidth,angle=0]{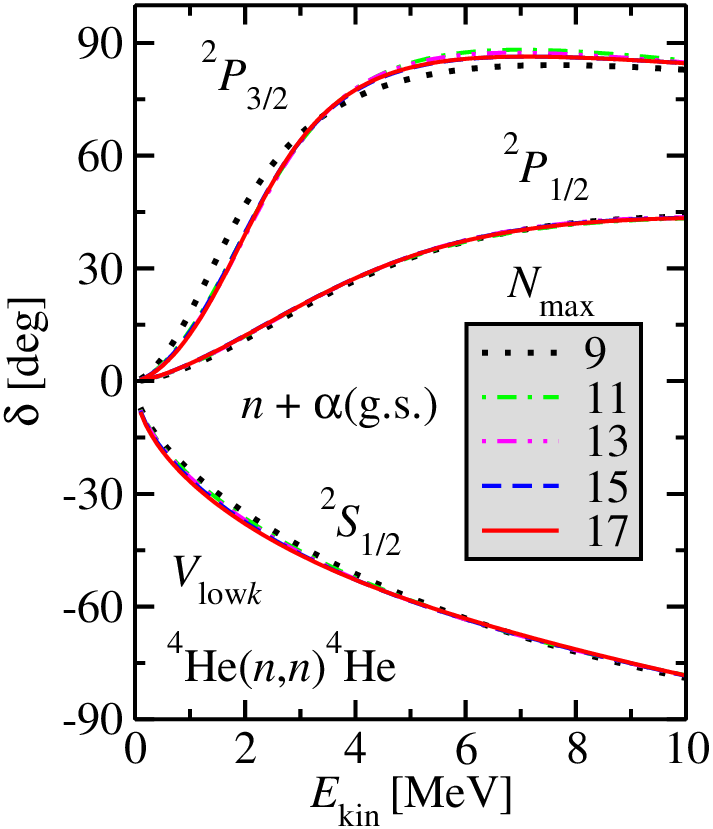}
\caption{Dependence on $N_{\rm max}$ of the $n$-$\alpha({\rm g.s.})$ phase shifts with the $V_{{\rm low}k}$~\cite{QuaglioniS_Vlowk} $NN$ potential at $\hbar\Omega=18$ MeV.}
\label{QuaglioniS_fig:4}       
\end{figure}
\begin{figure}[!htb]
\centering
\includegraphics[width=0.8\columnwidth,angle=0]{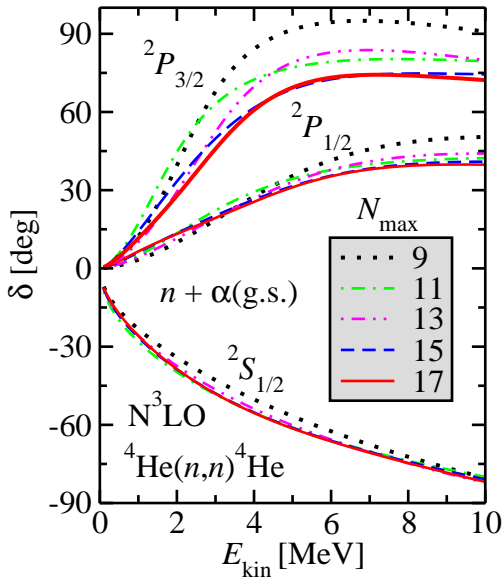}
\caption{Dependence on $N_{\rm max}$ of the $n$-$\alpha({\rm g.s.})$ phase shifts with the N$^3$LO~\cite{QuaglioniS_N3LO} $NN$ potential at $\hbar\Omega=19$ MeV.}
\label{QuaglioniS_fig:5}       
\end{figure}
\begin{figure}[!htb]
\centering
\includegraphics[width=0.8\columnwidth,angle=0]{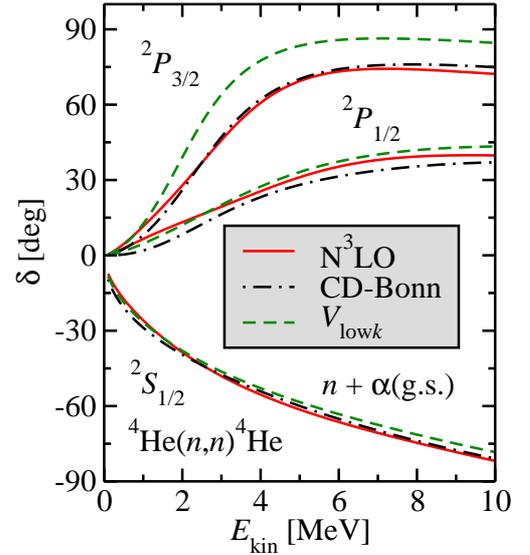}
\caption{Calculated $n$-$\alpha({\rm g.s.})$ phase shifts obtained in the largest model space ($N_{\rm max}=17$) for the $V_{{\rm low}k}$~\cite{QuaglioniS_Vlowk}  $NN$ potential at $\hbar\Omega=18$ MeV and the N$^3$LO~\cite{QuaglioniS_N3LO} and  CD-Bonn~\cite{QuaglioniS_cdb2k} $NN$ interactions at $\hbar\Omega=19$ MeV. }
\label{QuaglioniS_fig:6}       
\end{figure}
The two-cluster NCSM/RGM  formalism within the single-nucleon projectile outlined in the previous section, can be used to calculate nucleon-nucleus phase shifts below three-body break threshold, by solving the 
system of multi-\\channel Schr{\"o}dinger equations~(\ref{r-matrix-eq}) with scattering boundary conditions.  In the next sections we will review part of the results for neutrons scattering on $^3$H, $^4$He and $^{10}$Be and protons scattering on $^{3,4}$He, using realistic $NN$ potentials, which were first presented in Refs.~\cite{QuaglioniS_NCSMRGM_letter} and~\cite{QuaglioniS_NCSMRGM_prc}, and present some new calculations.   

\subsection{Convergence with respect to the HO model space}
\label{QuaglioniS_convergence}
To study the behavior of our approach with respect to the HO model space, we have performed NCSM/RGM scattering calculations for the $A=5$ systems, using the $V_{{\rm low}k}$ $NN$ potential~\cite{QuaglioniS_Vlowk}, which is ``soft" and we treated as ``bare", and the N$^3$LO $NN$ interaction~\cite{QuaglioniS_N3LO}, which generates strong short-range correlations, thus requiring the use of effective interactions. In particular, for this convergence tests, we restricted our binary-cluster basis to target-nucleon channel states with the target in its g.s.\ (corresponding to channel indexes of the type $\nu=\{4\;{\rm g.s.}\,0^+ 0;\,1\frac12^+\frac12;\,\frac12\,\ell\}$). 

Results obtained for $V_{{\rm low}k}$ are presented in Fig.~\ref{QuaglioniS_fig:4}.  The overall convergence is quite satisfactory, with a weak dependence on $N_{\rm max}$.  

Figure~\ref{QuaglioniS_fig:5} presents the convergence rate (achieved by using two-body effective interactions tailored to the HO model-space truncation) obtained for the same $n$-$\alpha$ scattering phase shifts with the N$^3$LO potential. Clearly, the N$^3$LO results  converge at a much slower rate than the $V_{{\rm low}k}$ ones. However, a gradual suppression of the difference between adjacent $N_{\rm max}$ values with increasing model-space size is visible, although the pattern is somewhat irregular for the $P$ phase shifts. 

Although not shown, the $p$-$\alpha$ phase shifts present analogous convergence properties. 

The next figure, Fig.~\ref{QuaglioniS_fig:6} compares the $N_{\rm max}\!=\!17$ results for the previously discussed $V_{{\rm low}k}$ and N$^3$LO $NN$ interactions, and those obtained with the CD-Bonn $NN$ potential~\cite{QuaglioniS_cdb2k}. The NCSM/RGM calculations for the latter potential were carried out using two-body effective interactions, and present a convergence pattern similar  to the one observed for N$^3$LO. Clearly, the $^2P_{1/2}$ and $^2P_{3/2}$ phase shifts are sensitive to the interaction models, and, in particular, to the strength of the spin-orbit force. This observation is in agreement with what was found in the earlier study of Ref.~\cite{QuaglioniS_GFMC_nHe4}.   
Following a behavior already observed in the structure of $p$-shell nuclei, CD-Bonn and N$^3$LO interactions yield about the same spin-orbit splitting. On the contrary, the larger separation between the $V_{{\rm low}k}$ $3/2^-$ and $1/2^-$ resonant phase shifts is direct evidence for a stronger spin-orbit interaction. 
\begin{figure*}[!htb]
\centering
\includegraphics[width=1.5\columnwidth,angle=0]{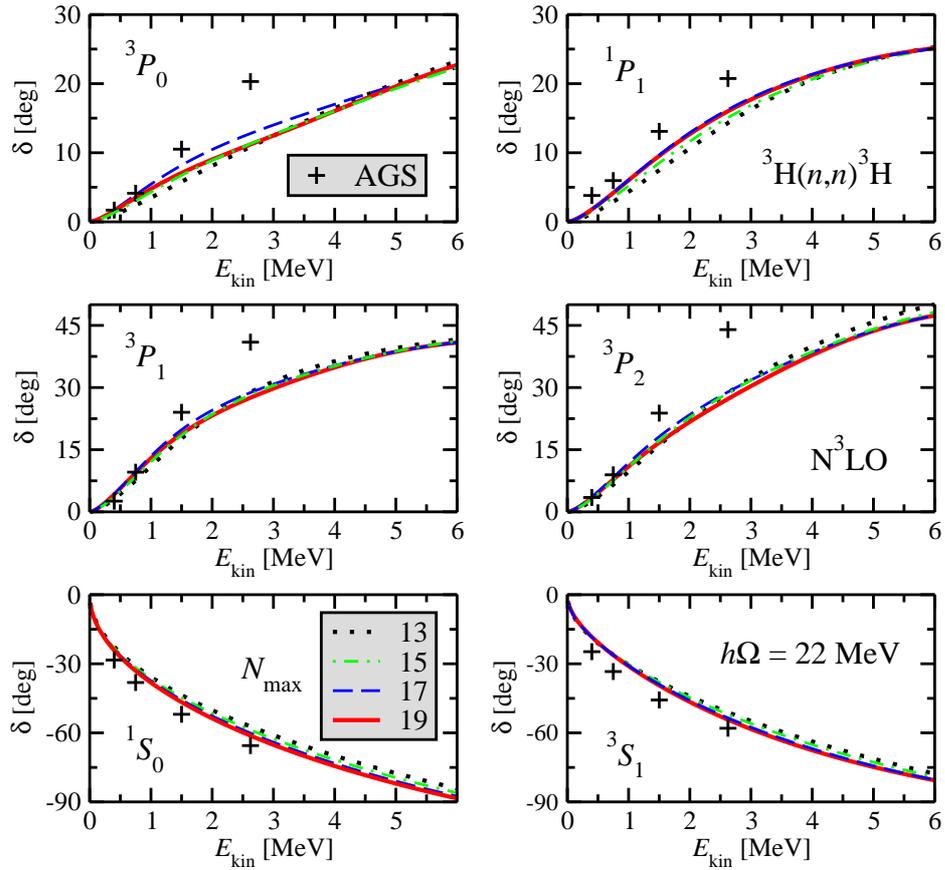}
\caption{Calculated 
 $n\,$-${}^3$H phase shifts as a function of the relative kinetic energy in the c.m.\ frame $E_{\rm kin}$, using the N$^3$LO $NN$ potential~\cite{QuaglioniS_N3LO} in the model spaces $N_{\rm max}=11-19$, at $\hbar\Omega = 22$ MeV. All results were obtained in a coupled-channel calculation including only the g.s.\ of the ${}^3$H nucleus  (i.e. the channels $\nu=\{3\,{\rm g.s.}\,\frac12^+\frac12;\,1\frac12^+\frac12;\,s\,\ell\}$).}
\label{QuaglioniS_fig:7}       
\end{figure*}
\begin{figure*}[!htb]
\centering
\includegraphics[width=1.5\columnwidth,angle=0]{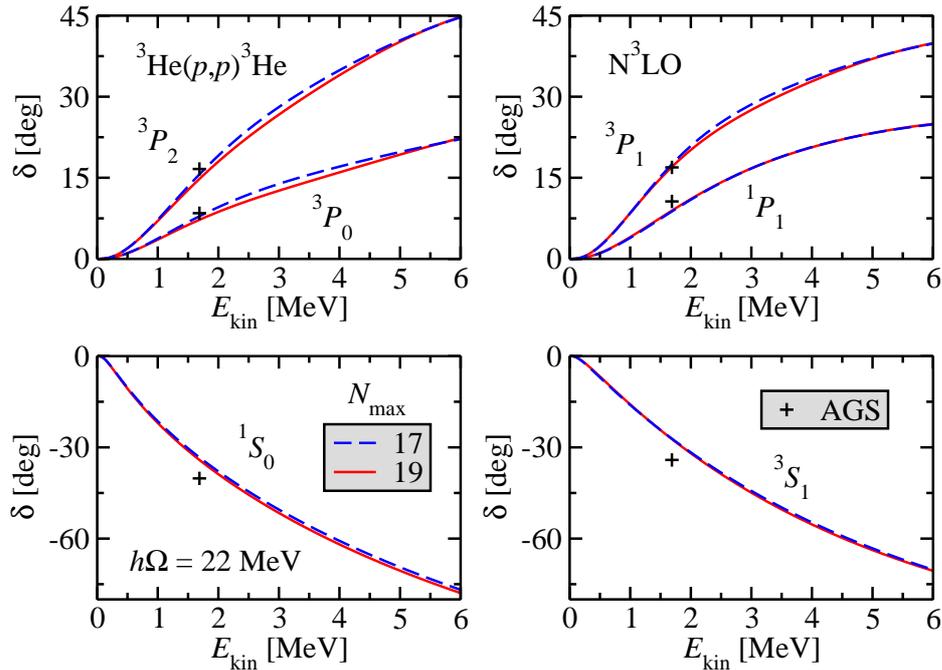}
\caption{Calculated 
 $p\,$-${}^3$He phase shifts for the N$^3$LO $NN$ potential~\cite{QuaglioniS_N3LO} in the model spaces $N_{\rm max}=17-19$, at $\hbar\Omega = 22$ MeV, compared to AGS results of Ref.~\cite{QuaglioniS_DeltuvaPriv}. All NCSM/RGM results were obtained in a coupled-channel calculation including only the g.s.\ of the ${}^3$He nucleus  (i.e. the channels $\nu=\{3\,{\rm g.s.}\,\frac12^+\frac12;\,1\frac12^+\frac12;\,s\,\ell\}$).}
\label{QuaglioniS_fig:8}       
\end{figure*}

As the $1/2^+$ channel is dominated by the repulsion between the neutron and the $\alpha$ particle induced by the Pauli exclusion principle (see also Figs.~\ref{QuaglioniS_fig:2}, \ref{QuaglioniS_fig:3}), the short-range details of the nuclear interaction play a minor role on the $^2S_{1/2}$ phase shifts. As a consequence, we find very similar results for all of the three adopted $NN$ potential models. 

\subsection{Test of the NCSM/RGM approach in the four-nucleon sector}
\label{QuaglioniS_A4}
A stringent test-ground to study the performances of the NCSM/RGM approach within the single-nucleon projectile basis is provided by the four-nucleon system. Numerically exact calculations for the $A=4$ sector have been already successfully performed within accurate few-body techniques, such as  Faddeev-Yacubovsky, AGS, and hyperspherical harmonics methods~\cite{QuaglioniS_Lazauskas05-1,QuaglioniS_Lazauskas09,QuaglioniS_Deltuva07-1,QuaglioniS_Deltuva07-2,QuaglioniS_Viviani09,QuaglioniS_Lazauskas05-2}.

Figures~\ref{QuaglioniS_fig:7} and~\ref{QuaglioniS_fig:8} show $n\,$-${}^3$H and $p\,$-${}^3$He phase shifts, respectively, calculated with the N$^3$LO $NN$ potential~\cite{QuaglioniS_N3LO} together with results obtained by Deltuva and Fonseca~\cite{QuaglioniS_Deltuva07-1,QuaglioniS_DeltuvaPriv} from the solution of the AGS equations ($+$ symbols) using the same interaction. 
The convergence behavior of The NCSM/RGM calculations was achieved using two-body effective interactions tailored to the model-space truncation, as outlined in Sec.~\ref{QuaglioniS_ncsmrgm}. For the $^1S_{0}$, $^1P_1$ and $^3S_1$ partial waves, the increase in model-space size produces gradually smaller deviations with a clear convergence towards the $N_{\rm max}=19$ results.  The rest of the phase shifts, particularly the $^3P_0$, show a more irregular pattern. Nevertheless, in the whole energy-range we find less than $2$ deg absolute difference between the phases obtained in the largest and next-to-largest model spaces.
 
Concerning the comparison to the highly accurate AGS results,  in general  the agreement between the two calculations worsens as the relative kinetic energy in the c.m.\ frame, $E_{\rm kin}$,  increases. This discrepancy is a manifestation of the the influence played by closed channels not included in our basis states, that is, target-nucleon channel states with the target above the $I_1^{\pi_1}=\frac12^+$ g.s., and 2+2 configurations, both of which are taken into account by the AGS results. 
Because these states correspond to the breakup of the $A=3$ system, it is not feasible to include them in the current version of the NCSM/RGM approach, which so far has been derived only in the single-nucleon projectile basis. However, we are planning on extending our approach to be able to account for the target breakup, and these development will be discussed in future publications. Nevertheless, this test of the NCSM/RGM approach in the $A=4$ sector clearly indicates that one has to pay attention not only to the convergence with respect to the HO model-space size $N_{\rm max}$, but also to the convergence in the RGM model space, which is enlarged by including excited states of the nucleon clusters in the binary-channel basis states.

\subsection{$^{\bf 4}$He$\boldsymbol{(N,N)^4}$He scattering}
A better scenario for the application of the NCSM/RGM approach within the single-nucleon projectile basis is the scattering of nucleons on $^4$He. This process is characterized by a single open channel up to the $^4$He breakup threshold, which is fairly high in energy. In addition the low-lying resonances of $^4$He are narrow enough that they can be reasonably reproduced by diagonalizing the four-body Hamiltonian in the NCSM model space, and consistently included as closed channels in the NCSM/RGM model\\ space. 

\begin{figure}[!htb]
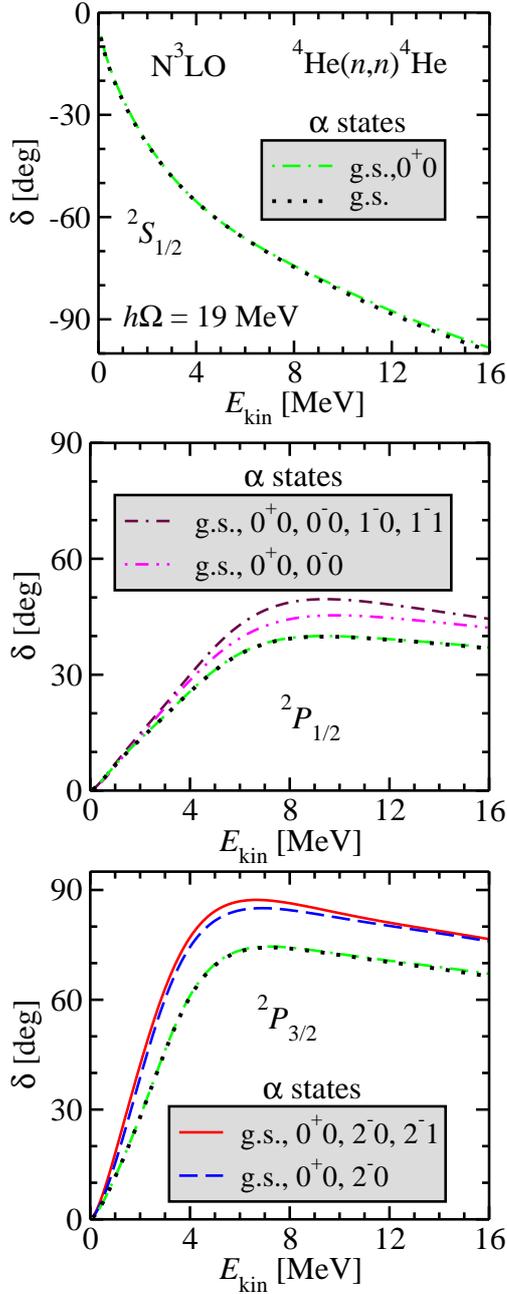

\centering
\includegraphics[width=0.8\columnwidth,angle=0]{QuaglioniS-fig9a.eps}
\includegraphics[width=0.8\columnwidth,angle=0]{QuaglioniS-fig9b.eps}
\includegraphics[width=0.8\columnwidth,angle=0]{QuaglioniS-fig9c.eps}
\caption{Influence of  the lowest six excited states ($I_1^{\pi_1}T_1 = 0^+ 0, 0^- 0,1^-0,1^-1, 2^- 0,2^-1$) of the $\alpha$ particle on the $n$-$\alpha$ $^2S_{1/2}$ (top panel), $^2P_{1/2}$ (central panel), and $^2P_{3/2}$ (bottom panel) phase-shift results for the N$^3$LO $NN$ potential~\cite{QuaglioniS_N3LO} at $\hbar\Omega=19$ MeV. Dotted (g.s.) and dash-dotted (g.s., $0^+0$) lines correspond to single- and coupled-channel calculations in a $N_{\rm max} = 17$ model space, respectively. The effects on the $^2P_{1/2}$ and $^2P_{3/2}$ phase shifts of the further inclusion of, respectively, the $0^- 0,1^-0,1^-1$, and $2^- 0,2^-1$ states are investigated in a $N_{\rm max}=15$ model space. }
\label{QuaglioniS_fig:9}       
\end{figure}

In Fig.~\ref{QuaglioniS_fig:9} we explore the effect of the inclusion of the first six excited states of the $^4$He on the $n$-$\alpha$ scattering phase shifts obtained with the N$^3$LO $NN$ interaction. 
More specifically, 
in addition to the single-channel results (dotted line) discussed in Sec.~\ref{QuaglioniS_convergence}, we show coupled-channel calculations for five different combinations of $^4$He states, i.e., $i)$ g.s.,$0^+0$ (dash-dotted line), $ii)$ g.s.,$0^+0,0^-0$ (dash-dot-dotted line), $iii)$ g.s.,$0^+0,0^-0,1^-0,1^-1$ (dash-dash-dotted line),  $iv)$ g.s.,$0^+0,2^-0$ (dashed line), and $v)$ g.s.,$0^+0,2^-0,2^-1$ (solid line). 

The use of these five different combinations of ground and excited states (also shown in the legends of Fig.~\ref{QuaglioniS_fig:9}) indicates that the $^2S_{1/2}$ phase shifts are well described already by coupled channel calculations with g.s. and first $0^+0$ (the $^2S_{1/2}$ phase shifts obtained in the four larger Hilbert spaces are omitted for clarity of the figure). On the contrary, the negative parity excited states have relatively large effects on the $P$ phase shifts,  and in particular the $0^-0,1^-0$ and $1^-1$ mostly on the $^2P_{1/2}$, whereas the $2^-0$ and $2^-1$ on the $^2P_{3/2}$.
These negative parity states influence the $P$ phase shifts because they introduce couplings to the $s$-wave of relative motion.  Though also $I_1^{\pi_1}\!=\!1^-$ couples to  $\ell=0$ in the $3/2^-$ channel, the coupling of the $I_1^{\pi_1}=2^-$ states is dominant for the $^2P_{3/2}$ phase shifts. 
\begin{figure}[!htb]
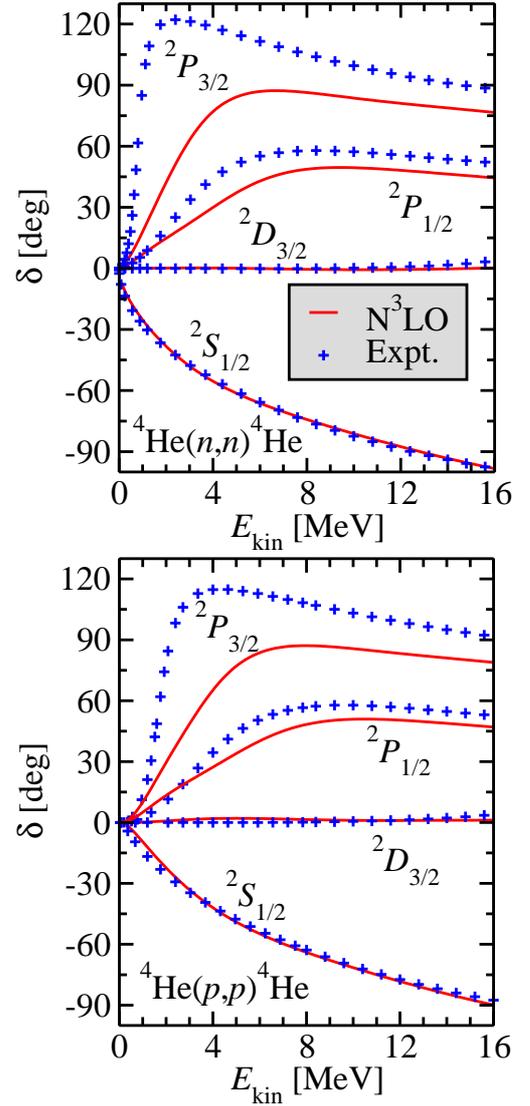

\centering
\includegraphics[width=0.8\columnwidth,angle=0]{QuaglioniS-fig10a.eps}
\includegraphics[width=0.8\columnwidth,angle=0]{QuaglioniS-fig10b.eps}
\caption{Calculated $n$- (top panel) and $p\,$-$\alpha$ (bottom panel) phase shifts for the N$^3$LO $NN$ potential~\cite{QuaglioniS_N3LO}  
compared to an $R$-matrix analysis of data ($+$)~\cite{QuaglioniS_HalePriv}. The $^2S_{1/2}$, $^2P_{1/2}$ and $^2P_{3/2}$ phase shifts correspond to the dotted(g.s., $0^+0$), dash-dash-dotted (g.s.,$0^+0,0^-0,1^-0,1^-1$), and solid (g.s.,$0^+0,2^-0,2^-1$) lines, respectively, of Fig.~\ref{QuaglioniS_fig:9}. The $^{2}D_{3/2}$ phase shifts were obtained in a coupled-channel calculation including ground and first $0^+$ excited state of $^4$He, in a $N_{\rm max}=17$ HO model space. }
\label{QuaglioniS_fig:10}       
\end{figure}

Figure~\ref{QuaglioniS_fig:10} compares an accurate $R$-matrix analysis of the nucleon-$\alpha$ scattering~\cite{QuaglioniS_HalePriv} with NCSM/RGM results obtained including the first six $^4$He excited states as in Fig.~\ref{QuaglioniS_fig:9}. This comparison reveals that for both neutron (top panel) and proton (bottom panel) projectiles we can describe quite well the $^2S_{1/2}$ and qualitatively also the $^2D_{3/2}$ phase shifts, using the N$^3$LO $NN$ potential. On the contrary, the same interaction is not able to reproduce well the two $P$ phase shifts, which are both too small and too close to each other. This lack of spin-orbit splitting between the $^2P_{1/2}$ and $^2P_{1/2}$ results can be explained by the omission in our treatment of the $NNN$ terms of the chiral interaction, which would provide additional spin-orbit force. This sensitivity of the $P$ phases to the strength of the spin-orbit force corroborated by the differences among the $V_{{\rm low}k}$, N$^3$LO and CD-Bonn results in Fig.~\ref{QuaglioniS_fig:4}:  $^2P_{1/2}$  and $^2P_{3/2}$ are both larger and more apart for $V_{{\rm low} k}$ than for N$^3$LO or CD-Bonn potentials.  At the same time, the good agreement of the N$^3$LO  and CD-Bonn $^2S_{1/2}$ phase shifts with their $V_{{\rm low}k}$ analogous and the $R$-matrix analysis owes to the repulsive action (in this channel) of the Pauli exclusion principle for short nucleon-$\alpha$ distances, which has the effect to mask the short-range details of the nuclear interaction. 

\begin{figure}[!htb]
\centering
\includegraphics[width=0.8\columnwidth,angle=0]{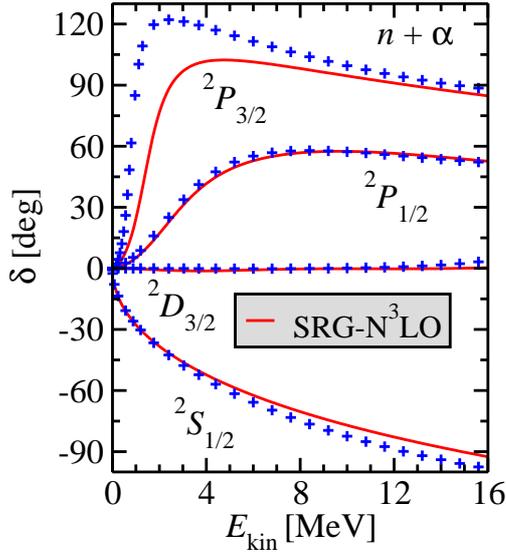}
\caption{Calculated $n\,$-$\alpha$ phase shifts obtained in a  $N_{\rm max}=17$ model space, including g.s. and $0^+0$ states of $^4$He, for the SRG-N$^3$LO~\cite{QuaglioniS_Roth_SRG}  $NN$ potential at $\hbar\Omega=20$ MeV, compared to an $R$-matrix analysis of data ($+$)~\cite{QuaglioniS_HalePriv}.}
\label{QuaglioniS_fig:11}       
\end{figure}
\begin{figure}[!hb]
\centering
\includegraphics[width=0.85\columnwidth,angle=0]{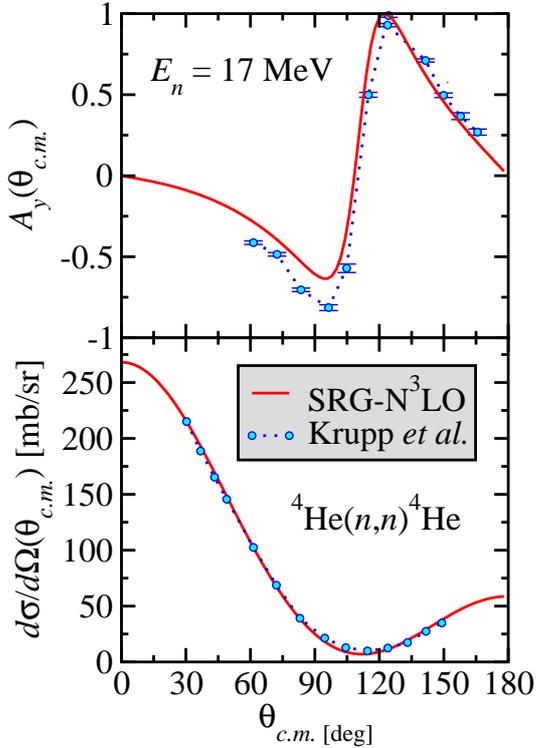}
\caption{Calculated $^4$He$(n,n)^4$He analyzing power (top panel) and differential cross section (bottom panel) for $E_n = 17$ MeV (neutron laboratory energy) obtained in the model space $N_{\rm max}=17$, including g.s. and $0^+0$ states of $^4$He, for the SRG-N$^3$LO~\cite{QuaglioniS_Roth_SRG}  $NN$ potential at $\hbar\Omega=20$ MeV, compared to the experimental data of Ref.~\cite{QuaglioniS_Krupp}.}
\label{QuaglioniS_fig:12}       
\end{figure}

As shown in Fig.~\ref{QuaglioniS_fig:11}, a stronger  spin-orbit interaction is shown also by the $^4$He$(n,n)^4$He $3/2^-$ and $1/2^-$ resonant phase shifts obtained using the SRG-evolved N$^3$LO $NN$ potential with cutoff $\Lambda=2.02$ fm$^{-1}$~\cite{QuaglioniS_Roth_SRG}, in a NCSM/\\RGM $n$-$\alpha$ model space including g.s. and $0^+0$ states of $^4$He. As for $V_{{\rm low}k}$ the two resonances present larger separation. In addition,  in the SRG-N$^3$LO case, the $^2P_{1/2}$ phase shifts lie on top of the $n$-$\alpha$ $R$-matrix analysis of Ref.~\cite{QuaglioniS_HalePriv} in the whole energy range, and one finds a satisfactory agreement with the $R$-matrix results for all four lowest partial waves starting from about $E_{\rm kin}=12$ MeV. This particularly good quality of the SRG-N$^3$LO $N$-$^4$He phase shifts for energies far from the low-lying resonances is the reason behind the the fairly good agreement, presented in Fig.~\ref{QuaglioniS_fig:12}, of the calculated $^4$He$(n,n)^4$He analyzing power (top panel) and differential cross section (bottom panel) with the experimental data of the $17$ MeV polarized-neutron experiments of Ref.~\cite{QuaglioniS_Krupp}.

\subsection{$\boldsymbol{n}-^{\bf 10}$Be scattering and $^{\bf 11}$Be bound states}
\label{QuaglioniS_A11}
\begin{figure}[!ht]
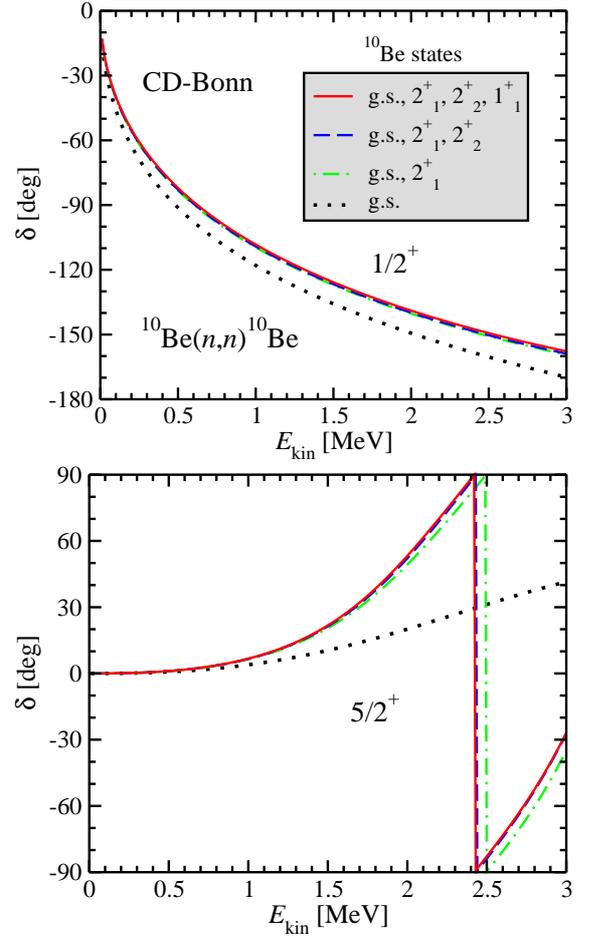

\centering
\includegraphics[width=0.9\columnwidth,angle=0]{QuaglioniS-fig13a.eps}
\includegraphics[width=0.9\columnwidth,angle=0]{QuaglioniS-fig13b.eps}
\caption{Calculated $n\,$-${}^{10}$Be phase shifts as a function of the relative kinetic energy in the c.m.\ frame $E_{\rm kin}$, using the CD-Bonn $NN$ potential~\cite{QuaglioniS_cdb2k} at $\hbar\Omega=13$ MeV: (top panel) $^2S_{1/2}$ and (bottom panel) $^2D_{5/2}$ results. The NCSM/RGM results were obtained using $n+^{10}$Be configurations with $N_{\rm max}$ = 6 g.s., $2^+_1$, $2^+_2$, and $1^+_1$ states of $^{10}$Be. The obtained $^2S_{1/2}$ scattering length is $+10.7$ fm.}
\label{QuaglioniS_fig:13}       
\end{figure}

Figure~\ref{QuaglioniS_fig:13} highlights one of the promising aspects of the NCSM/RGM approach, that is the ability (through the use of SD channel states) to perform {\em ab initio} scattering calculations for $p$-shell nuclei. The $^2S_{1/2}$ (top panel) and $^2D_{5/2}$ (bottom panel) $n$-$^{10}Be$ phase shifts  were obtained in a $N_{\rm max}=6$, $\hbar\Omega=13$ HO model space. The inclusion of the $2_1^+$ excited state of $^{10}$Be has a significant effect on the $S$ and more importantly on the $D$ phase, where it is essential for the appearance of a resonance below 3 MeV.  We note that a resonance has been observed at $\sim 1.8$ MeV with a tentative spin assignment  of $(5/2,3/2)^+$~\cite{QuaglioniS_AS90}. The further addition of the $2_2^+$ and especially $1_1^+$ excited states produces rather weak differences. We have also extracted the scattering length for the $^2S_{1/2}$ partial wave, and found a result of $+10.7$fm, which is comparable to the value of $+13.6$ fm obtained by Descouvemont in Ref.~\cite{QuaglioniS_11Be-Volkov}, by fitting the experimental binding energy of $^{11}$Be. 

Although we mainly described its scattering applications, the NCSM/RGM is a powerful tool also for structure calculations, particularly for loosely-bound systems. By imposing bound-state boundary conditions to the set of coupled channel Schr\"odinger equations of Eq.~(\ref{r-matrix-eq}), we tested the performance of our single-nucleon projectile\\ NCSM/RGM formalism for the description of one-nucleon halo systems. In particular, because of the well-known  parity-inversion between its two bound states with respect to the predictions of the simple shell model~\cite{QuaglioniS_Talmi}, the $^{11}$Be nucleus represents an excellent test ground for our approach.

Large-scale {\em ab initio} NCSM calculations with several accurate $NN$ potentials of the $^{11}$Be low-lying spectrum were not able to explain its g.s. parity inversion~\cite{QuaglioniS_Fo05}. The explanation for these results can be searched into two main causes: (i) the size of the HO basis was not large enough to reproduce the correct asymptotic of the $n$-$^{10}$Be component of the 11-body wave function; (ii) the $NNN$ force, not included in the calculation, plays an important role in the inversion mechanism.  The second hypothesis was corroborated by the results obtained with the INOY $NN$ potential~\cite{QuaglioniS_INOY}, which seemed to indicate the possibility to reach the inversion in a large NCSM basis. 

By studying the $^{11}$Be bound states in a NCSM/RGM model space spanned by $n$-$^{10}$Be channel states with inclusion of the $N_{max}=6$ g.s., $2_1^+,2_2^+$, and $1_1^+$ of $^{10}$Be, we are now in the position to address the first hypothesis. Indeed, the correct asymptotic behavior of the $n$-$^{10}$Be wave functions is described naturally in the NCSM/RGM approach.
\begin{table*}
\caption{Calculated energies (in MeV) of the $^{10}$Be g.s.\ and  of the lowest negative- and positive-parity states in $^{11}$Be, obtained using the CD-Bonn $NN$ potential~\cite{QuaglioniS_cdb2k} at $\hbar\Omega=13$ MeV. The NCSM/RGM results were obtained using $n+^{10}$Be configurations with $N_{\rm max}$ = 6 g.s., $2^+_1$, $2^+_2$, and $1^+_1$ states of $^{10}$Be.}
\label{QuaglioniS_tab:1}       
\begin{tabular}{clccclrclrc}
\hline\noalign{\smallskip}
&&&$^{10}$Be&&\multicolumn{2}{c}{$^{11}$Be($\frac12 ^-$)}&&\multicolumn{2}{c}{$^{11}$Be($\frac12 ^+$)}&\\[0.7mm]\cline{4-4}\cline{6-7}\cline{9-10}\\[-4mm]
&&$N_{\rm max}$&$E_{\rm g.s.}$ &&\multicolumn{1}{c}{$E$}&\multicolumn{1}{c}{$E_{th}$}&&\multicolumn{1}{c}{$E$}&\multicolumn{1}{c}{$E_{th}$}&\\[0.5mm]
\noalign{\smallskip}\hline\noalign{\smallskip}
&NCSM~\cite{QuaglioniS_10Be,QuaglioniS_Fo05} & $8/9$& $-57.06$&&$-56.95$&$0.11$&&$-54.26$&$2.80$&\\
&NCSM~\cite{QuaglioniS_10Be,QuaglioniS_Fo05,QuaglioniS_present} & $6/7$& $-57.17$&&$-57.51$&$-0.34$&&$-54.39$&$2.78$&\\
&NCSM/RGM~\cite{QuaglioniS_present} &&&&$-57.59$&$-0.42$&&$-57.85$&$-0.68$&\\
&Expt. & & $-64.98$&&$-65.16$&$-0.18$&&$-65.48$&$-0.50$&\\ 
\noalign{\smallskip}\hline
\end{tabular}
\end{table*}
The energies of the lowest $1/2^+$ and $1/2^-$ states of $^{11}$Be obtained in the NCSM and in the NCSM/RGM calculations, using the same CD-Bonn NN interaction~\cite{QuaglioniS_cdb2k} at $\hbar\Omega=13$ MeV adopted in Ref.~\cite{QuaglioniS_Fo05}, are presented in Table~\ref{QuaglioniS_tab:1}.  The relatively small differences between the $N_{\rm max}=6/7$ and $N_{\rm max}=8/9$ NCSM results, seems to indicate a reasonable degree of convergence for these calculations. The $1/2^-$ state appears to be the g.s., and the $1/2^+$ state is about 2.8 MeV above the $n\,$-$^{10}$Be threshold. A comparison to the NCSM/RGM calculations (obtained in a model space including g.s., $2^+_1$, $2^+_2$, and $1^+_1$ states of $^{10}$Be) shows a rough agreement  for the $1/2^-$ state, whereas for the $1/2^+$ state one observes a dramatic difference ($\sim$3.5 MeV) in the energy. The $1/2^-$ and $1/2^+$ NCSM/RGM states are both bound and the $1/2^+$ state is the g.s.\ of $^{11}$Be. Correspondingly, we obtain a B(E1; $\frac12^-\rightarrow\frac12^+$) value of $0.18$ $e^2$ fm$^2$, which is not far from experiment.

\begin{table}[!b]
\caption{Mean values of the relative kinetic and potential energy and of the internal $^{10}$Be energy in the $^{11}$Be $1/2^+$ ground state. All energies in MeV. NCSM/RGM calculation as in Table~\ref{QuaglioniS_tab:1}. See the text for further details.}\label{QuaglioniS_tab:2}
\begin{tabular}{lcccc}
\hline\noalign{\smallskip}
NCSM/RGM & $\langle T_{\rm rel} \rangle$ & $\langle W\rangle$ & $E[^{10}{\rm Be(g.s.,ex.)}]$ & $E_{\rm tot}$\\
\noalign{\smallskip}\hline\noalign{\smallskip}
Model Space & $16.65$  & $-15.02$   & $-56.66$ & $-55.03$ \\
Full                  & $\;\,6.56$ & $\;\,-7.39$ & $-57.02$ & $-57.85$\\
\noalign{\smallskip}\hline
\end{tabular}
\end{table}
To understand the mechanism which makes the $1/2^+$ state bound in the NCSM/RGM, we evaluated mean values of the relative kinetic and potential energies as well as the mean value of the $^{10}$Be energy, and compared them to those obtained by restricting all the integration kernels to the HO model space (i.e. by replacing the delta function of Eq.~(\ref{norm}) with its representation in the HO model space). These results are shown in Table~\ref{QuaglioniS_tab:2}. The model-space-restricted calculation is then similar, although not identical, to the standard NCSM calculation. In particular, as in the NCSM one loses the correct asymptotic behavior of the $n$-$^{10}$Be wave function. We observe that in the full NCSM/RGM calculation both relative kinetic and potential energies are smaller in absolute value. This is an effect of the re-scaling of the relative wave function in the internal region when the Whittaker tail is recovered. The difference is significantly more substantial for the relative kinetic energy than for the potential energy. As a result one obtains a dramatic decrease of the energy of the $1/2^+$ state, which makes it bound and even leads to a g.s.\ parity inversion. This study shows that a proper treatment of the coupling to the $n\,$-${}^{10}$Be continuum is essential in explaining the g.s.\ parity inversion. However, we cannot exclude that the $NNN$ force plays a role in the inversion mechanism, not until accurate calculations with both $NNN$ force and full treatment of the $n$-$^{10}$Be tail will be performed.   

\section{Conclusions and Outlook}
\label{QuaglioniS_conclusions}

\begin{figure}[!b]
\centering
\includegraphics[width=0.9\columnwidth,angle=0]{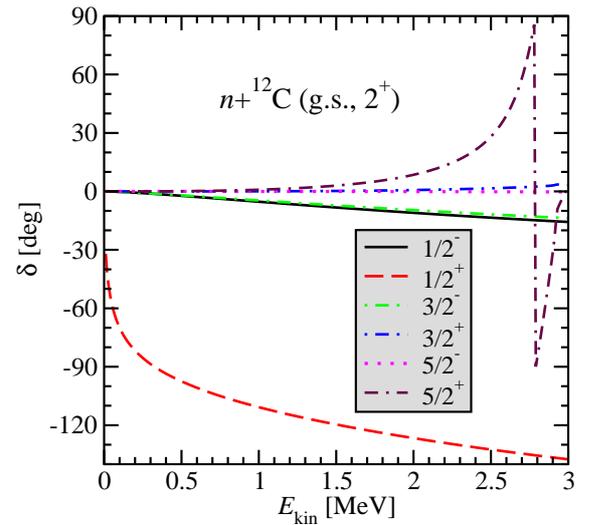}
\caption{Calculated $n-^{12}$C phase shifts obtained within the NCSM/RGM in a $N_{\rm max}=16$ model space including g.s. and first excited states of $^{12}$C, using the SRG-N$^3$LO NN potential of Ref.~\cite{QuaglioniS_Roth_SRG} with cutoff  $\Lambda = 2.66$ fm$^{-1}$ and HO frequency $\hbar\Omega=24$ MeV. The $^{12}$C wave functions were obtained within the importance-truncated NCSM~\cite{QuaglioniS_IT-NCSM,QuaglioniS_Roth09}.}
\label{QuaglioniS_fig:14}       
\end{figure}
\begin{figure*}[!htb]
\centering
\includegraphics[width=1.3\columnwidth,angle=0]{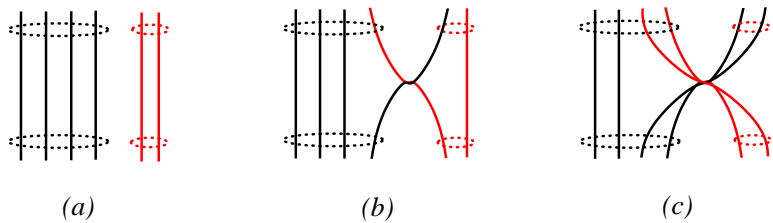}
\caption{Diagrammatic representation of: ($a$) ``direct", ($b$) ``single-exchange", and ($d$) ``double-exchange" components of the norm kernel for the $(A-2,2)$ cluster basis. The first group of circled lines represents the first cluster, the bound state of $A{-}2$ nucleons. The second couple of circled lines represents the second cluster, in the specific case a deuteron nucleus. Bottom and upper part of the diagram represent initial and final states, respectively. See also caption of fig.~\ref{QuaglioniS_fig:1}.}
\label{QuaglioniS_fig:15}       
\end{figure*}
We have reviewed the NCSM/RGM,  a new {\em ab initio} many-body approach capable 
of describing simultaneously both bound and scattering states in light nuclei,
by combining the RGM with the use of realistic interactions, and a microscopic and consistent description of the nucleon clusters, achieved via the {\em ab initio} NCSM.
In particular, we have outlined the formalism on which the NCSM/RGM is based, and given examples of the algebraic expressions for the integral kernels within the single-nucleon projectile basis, working both with Jacobi, and SD single-particle coordinate bases. As the spurious c.m.\ components present in the SD basis were removed exactly, in both frameworks the calculated integral kernels are translationally invariant, and lead to identical results. Several analytical as well as numerical tests were performed in order to verify the approach, particularly by benchmarking  independent Jacobi-coordinate and SD calculations for systems with up to 5 nucleons.
\begin{figure}[!b]
\centering
\includegraphics[width=0.9\columnwidth,angle=0]{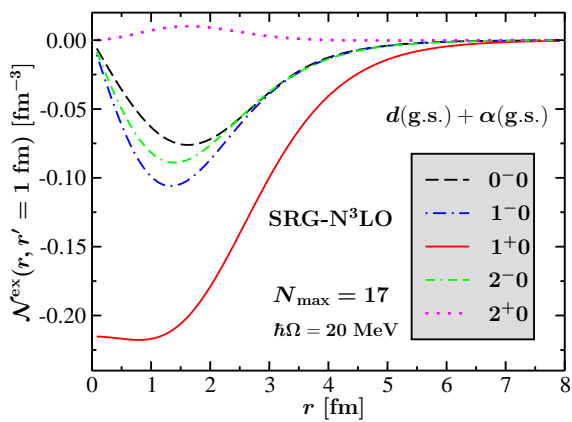}
\caption{Calculated ``exchange" part of the norm kernel, ${\mathcal N}^{J^\pi T}_{\ell\ell}(r^\prime,r) - \frac{\delta(r^\prime-r)}{r^\prime\,r}$ for the $d$(g.s.)$+\alpha$(g.s.) channels as a function of the relative coordinate $r$ at $r^\prime = 1$ fm, using the SRG-N$^3$LO $NN$ potential~\cite{QuaglioniS_Roth_SRG} at $\hbar\Omega=20$ MeV. Effects of the Pauli-exclusion principle are visible partcularly in the $1^+0$ channel.}
\label{QuaglioniS_fig:16}       
\end{figure}

Among the applications, we reviewed results for neutron scattering on $^3$H, $^4$He and $^{10}$Be and proton scattering on $^{3,4}$He, using realistic $NN$ potentials. Our $A=4$ scattering results were compared to earlier {\em ab initio} calculations performed in the framework of the AGS equations, and the convergence properties of the NCSM/RGM approach with respect to the adopted model spaces were discussed in detail. For the $A=5$ system, we found that all adopted  $NN$ potentials  provide a fairly good (in some case excellent) description of the $S$-wave phase shifts. On the contrary, the $P$-wave phase shifts that we obtained with any of the realistic $NN$ potentials present both insufficient magnitude and splitting with respect to the $R$-matrix analysis of the data. Concerning the calculations performed with the N$^3$LO $NN$ potential, it is anticipated that the inclusion of the $NNN$ terms of the chiral interaction would lead to an enhanced spin-orbit splitting, and recover the predictions of the $R$-matrix analysis. We also showed that the SRG-N$^3$LO interaction presents a larger spin-orbit strength than the N$^3$LO $NN$ potential itself, and allows a fairly good description of $^4$He$(n,n)^4$He angular cross section and analyzing power for energies far above the low-lying resonances. 
An important aspect of the NCSM/RGM approach is its suitability for the description of loosely-bound systems, such as the $^{11}$Be nucleus. Although we cannot exclude that, e.g. the $NNN$ force plays a role in the inversion mechanism, we have demonstrated that a proper treatment of the coupling to the $n\,$-${}^{10}$Be continuum leads to a dramatic decrease of the energy of the $\frac12^+$ state, which makes it bound and even leads to a g.s.\ parity inversion.

Since the publication of the first results~\cite{QuaglioniS_NCSMRGM_letter,QuaglioniS_NCSMRGM_prc}, the NSCM/RGM approach has been applied to the description of nucleon scattering on several $p$-shell nuclei, such as $^7$Li, $^{12}$C, and $^{16}$O (see, {\em e.g.}, Fig.~\ref{QuaglioniS_fig:14}). Key to these calculations, which will be published in a forthcoming paper~\cite{QuaglioniS_NCSMRGM_topublish}, are two factors: $(i)$ the ability of the NCSM/RGM to take advantage of the powerful second quantization techniques, while preserving the  translational-invariance symmetry of the system; and $(ii)$ the use of large HO model spaces (large $N_{\rm max}$ values) for the expansion of the clusters internal wave functions, and hence of the short-range parts of the integration kernels. The use of the so-called importance-truncated NCSM~\cite{QuaglioniS_IT-NCSM} for the description of the clusters internal wave functions is essential in achieving the second of these two points. This method, first introduced by R. Roth and P. Navr\'atil in 2007~\cite{QuaglioniS_IT-NCSM} for the calculation of ground states, then further developed in Ref.~\cite{QuaglioniS_Roth09}, and now extended to the calculation of excited states~\cite{QuaglioniS_NCSMRGM_topublish}, allows to reach large HO model spaces for $p$-shell nuclei by means of an a-priori selection of the most important NCSM basis states. Using importance-truncated clusters wave functions within the NCSM/RGM formalism allows us to reach for scattering calculations on heavier nuclei the same level of convergence obtained here for the $A=4$ and $A=5$ systems.

The NCSM/RGM formalism has been also extended to include two-nucleon (deuteron) projectiles, and the inclusion of the three-nucleon (triton and $^3$He) and four-nucleon ($^4$He) projectiles are planned ahead. 

Calculations of deuteron-nucleus scattering are underway. As an example, Fig.~\ref{QuaglioniS_fig:15} shows the diagrammatic representation of the matrix elements involved in the calculation of the norm kernel within the deuteron-projectile basis, and the ``exchange" component for such kernel for the $d$(g.s.)$+\alpha$(g.s.) system is presented in Fig.~\ref{QuaglioniS_fig:16}.   

In addition, the coupling of the single-nucleon and two-nucleon projectile basis will allow the first {\em ab initio} calculation of the $^3$H$(d,n)^4$He fusion. 

Further, it is possible and desirable to extend the binary-cluster $(A{-}a,a)$ NCSM/RGM basis by the standard $A$-\\nucleon NCSM basis to unify the original {\em ab initio} NCSM and NCSM/RGM approaches. This will lead to a much faster convergence of the many-body calculations compared to the original approaches and, most importantly, to an optimal and balanced unified description of both bound and unbound states. Extensions of the approach to include three-body cluster channels are also among our future plans, and the feasibility of such a project is supported by recent developments on the treatment of both three-body bound and continuum states (see, e.g., Refs.~\cite{QuaglioniS_GCM3bcont,QuaglioniS_3bbound1,QuaglioniS_3bcont1,QuaglioniS_3bcnfr,QuaglioniS_3bbound2}).   

\section{Acknowledgments}
\label{QuaglioniS_ack}
Numerical calculations have been performed in part at the LLNL LC facilities.
Prepared in part by LLNL under Contract DE-AC52-07NA27344.
S.Q. and P.N. acknowledge support from the U.\ S.\ DOE/SC/NP (Work Proposal No.\ SCW0498), LLNL LDRD grant PLS-09-ERD-020,
and from the U.\ S.\ Department of Energy Grant DE-FC02-07ER41457. R.R. acknowledges support from the DFG (SFB 634) and  
HIC for FAIR.

\section{Bibliography}
\label{QuaglioniS_bib}

\end{document}